\documentclass[showpacs,preprintnumbers,amsmath,amssymb,superscriptaddress]{revtex4}
\usepackage{bm}
\usepackage{graphicx}

\begin{document}

\title{Chiral polymerization: symmetry breaking and entropy production in closed systems}

\author{Celia Blanco}
\email{blancodtc@inta.es} \affiliation{Centro de Astrobiolog\'{\i}a
(CSIC-INTA), Carretera Ajalvir Kil\'{o}metro 4, 28850 Torrej\'{o}n
de Ardoz, Madrid, Spain}
\author{David Hochberg}
\email{hochbergd@inta.es} \affiliation{Centro de Astrobiolog\'{\i}a
(CSIC-INTA), Carretera Ajalvir Kil\'{o}metro 4, 28850 Torrej\'{o}n
de Ardoz, Madrid, Spain}

\begin{abstract}
We solve numerically a kinetic model of chiral polymerization in systems
closed to matter and energy flow, paying special emphasis to its
ability to amplify the small initial enantiomeric excesses due
to the internal and unavoidable statistical fluctuations. The
reaction steps are assumed to be reversible, implying a thermodynamic
constraint among some of the rate constants. Absolute asymmetric synthesis is
achieved in this scheme.  The system can persist for long times in quasi-stationary chiral
asymmetric states before racemizing. Strong inhibition leads to long-period chiral oscillations in the enantiomeric excesses of the
longest homopolymer chains. We also calculate the
entropy production $\sigma$ per unit volume  and show that $\sigma$
increases to a peak value either before or in the vicinity of the chiral
symmetry breaking transition.
\end{abstract}

\date{\today}

\maketitle

\section{\label{sec:intro} Introduction}

There is a growing consensus that the homochirality of biological compounds is a
condition associated to life that should have emerged in the abiotic
stages of evolution through processes of spontaneous mirror symmetry breaking (SMSB). This could
have proceeded in a prebiotic stage, incorporating steps of increasing complexity thus leading to
chemical systems and enantioselective chemical networks \cite{AG,Weissbuch,Guijarro}.
An important issue is therefore to identify processes of chirality amplification
in chemical reactions. In this regard, a recent kinetic analysis of the Frank model in closed systems applied to the
Soai reaction \cite{CHMR} has taught us that in an actual chemical scenario,
reaction networks that exhibit SMSB are very sensitive to chiral inductions owing to the
presence of tiny initial enantiomeric excesses, as previously shown theoretically \cite{KondeAvet}.
The stochastic scenario implies the creation of chirality from intrinsic chiral fluctuations and its later transmission and
amplification. This can occur in far-from-equilibrium systems that undergo dynamic phase transitions.

The process must be coupled to others which preserve, extend, and
transmit the chirality. Biological homochirality of living systems involves large
macromolecules, therefore a central point is the relationship of the polymerization process
with the emergence of chirality. This hypothesis has inspired recent activity devoted to modeling
efforts aimed at understanding mirror symmetry breaking in
polymerization of relevance to the origin of life.  The models
so proposed \cite{Sandars,BM,BAHN,WC,Gleisera,Gleiserb,Gleiserc,Gleiserd,SH}
are by and large, elaborate extensions and generalizations of
Frank's original paradigmatic scheme \cite{Frank}. Heading this list, Sandars
\cite{Sandars} introduced a detailed polymerization process plus the
basic elements of enantiomeric cross inhibition as well as a chiral
feedback mechanism in which only the largest polymers formed can enhance
the production of the monomers from an achiral substrate. He treated basic numerical studies of
symmetry breaking and bifurcation properties of this model for
various values of the number of repeat units $N$. All the subsequent
models cited here are variations on Sandars'
original theme. Soon afterwards, Brandenburg and coworkers \cite{BAHN} studied the
stability and conservation properties of a modified Sandars' model
and introduce a reduced $N=2$ version including the effects of
chiral bias. In \cite{BM}, they included spatial extent in this
model to study the spread and propagation of chiral domains
as well as the influence of a backround turbulent advection velocity
field. The model of Wattis and Coveney \cite{WC} differs from
Sandars' in that they allow for polymers to grow to arbitrary
lengths $N$ and the chiral polymers of all lengths, from the dimer and
upwards, act catalytically in the breakdown of the achiral source
into chiral monomers. An analytic linear stability analysis of both
the racemic and chiral solutions is carried out for the model's
large $N$ limit and various kinetic timescales are identified. The
role of external white noise on Sandars-type polymerization networks
including spatial extent has been explored by Gleiser and coworkers:
the $N=2$ truncated model introduced in \cite{BM} is subjected to
external white noise in \cite{Gleisera}, chiral bias is considered
in \cite{Gleiserb}, high intensity and long duration noise is
considered in \cite{Gleiserc} and in \cite{Gleiserd}, modified
Sandars-type models with spatial extent and external noise are
considered both for finite and infinite $N$, with an emphasis paid
to the dynamics of chiral symmetry breaking. By contrast, Saito and
Hyuga's \cite{SH} model gives rise to homochiral states but differs
markedly from Sandars' in that it does not invoke the enantiomeric
cross inhibition, allowing instead for reversibility in all the
reaction steps.  Their model requires open flow, which is the needed element of
irreversibility. A different model which stands apart from the above group is that due to Plasson et al. \cite{Plasson}.
They considered a recycled system based on reversible chemical reactions
and open only to energy flow and without any (auto)catalytic reactions.
A source of constant external energy --the element of irreversibility--is required to activate the monomers.
This energy could be introduced into the system in a physical form, say, as high energy photons. A
system of this kind, limited to dimerizations, was shown to have nonracemic stable final states for various
ranges of the model parameter values and for total concentrations greater than a minimal value.

The polymerization models referred to above are defined only for
\textit{open flow} systems which exchange matter and energy with
the exterior. A constant source of achiral precursor is usually
assumed. An unrealistic consequence is that homochiral chains can
grow to infinite length. By contrast, most experimental procedures
are carried out in \textit{closed and spatially bounded} reaction
domains and are initiated in far-from equilibrium states
\cite{Blocher,Weissbuch,Hitzb,Nery,Hitza,Rubinstein,Illos,Lahava,Lahavb}.
It is thus crucial to have models compatible with these experimentally
realistic boundary and initial conditions. The most immediate
consequences are that polymer chains can grow to a finite maximum
length and that the total system mass is constant.  These
two properties are of course intimately related. An original aspect of our
work is that we consider the polymerization process in a closed
system and include reversible reactions.
This enables us to explore absolute asymmetric synthesis in thermodynamically closed systems
(closed to matter flow) taking into account the backward reaction steps, which we call here
"reversible reactions/reversible models" in spite of the fact that the values of the forward and reverse rate constants
can be very different. As we are eloquently reminded by Mislow, mirror asymmetric
states are in practice unavoidable on purely statistical grounds alone \cite{Mislow}, even in the
absence of chiral physical forces. Absolute
asymmetric synthesis is the ability of a system to amplify these statistical and
minuscule enantiomeric biases up to observably large excesses. Thus a major goal of this work
is assess the ability of such polymerization schemes to amplify these initial excesses, albeit if only temporarily.
Asymmetric amplification is demonstrated to occur obeying microscopic reversibility in a
reversible model closed to matter flow.
This is of great practical interest as the racemization time scale
can be substantially longer than that of the transformation of the initial
reagents, depending on the strength of the mutual
inhibition, or direct interaction between the enantiomers. And regarding chemical evolution, this obviates the
need to invoke chiral physical fields and lends further support to the conviction that homochirality is a
``stereochemical imperative" of molecular evolution \cite{Siegel}.

The behavior of entropy in polymerization models is rarely
discussed \cite{Stier}, and has not been addressed for mirror symmetry breaking in chiral polymerization.
The entropy produced in a chemical reaction initiated out of equilibrium gives a measure
of the dissipation during the approach to final equilibrium. In this paper, we calculate the rate of
entropy production in chiral polymerization. Depending on the enantiomeric mutual inhibition, the entropy production
undergoes a rapid increase to a peak value either before, or else in the vicinity of the chiral
symmetry breaking transition, followed by an equally dramatic
decrease.  The system racemizes at
time scales greater than that of polymerization, and is accompanied by a final decrease of entropy
production to zero, indicating that the system has reached a final stable state
(not to be confused with a stationary state, which must be accompanied by a nonzero constant entropy production).
Computation of the average length homopolymer indicates that the final racemic state is dominated by the longest
available chains.

\section{\label{sec:model} The polymerization model}

The model we introduce and study here is modified and extended from that of Wattis and Coveney
\cite{WC} which is in turn, a generalization of Sandar's scheme \cite{Sandars}. Two salient differences
that distinguish our model from these and other previous ones \cite{Sandars,BM,BAHN,WC,Gleisera,Gleiserb,Gleiserc,Gleiserd,SH}
are that we (1) consider polymerization in \textit{closed} systems \cite{Hoch}--, so that
no matter flow is permitted with an external environment-- and (2) we allow for
reversible reactions in all the steps.
A third difference is that we
also include the formation (and dissociation) of the heterodimer.
While heterodimer formation was originally contemplated in
\cite{Sandars}, it has been silently omitted from all the subsequent
models \cite{BM,BAHN,WC,Gleisera,Gleiserb,Gleiserc,Gleiserd} that
derive therefrom. Fragmentation in a Sandars type model has been considered previously, but
was shown to yield a maximum average polymer length of only $N=3$ repeat units \cite{BAN}.

We assume there is an achiral precursor $S$ which can directly
produce the chiral monomers $L_1$ and $R_1$ at a slow rate
$\epsilon$ as well as be consumed in processes in which homopolymers
of all lengths catalyze the production of monomers. The specific
reaction scheme we study here is composed of the following steps,
where the $\epsilon$, ($\epsilon_{-}$), $k$ ($k_{-}$), etc., denote the forward (reverse)
reaction rate constants and $0 \leq f \leq 1$ is the fidelity of the feedback
mechanism:
\begin{equation}\label{precursor}
\begin{array}{ccc}
  S \stackrel{\epsilon}{\rightleftharpoons \atop{\small \epsilon_{-}}} L_1, &
  S + Q \stackrel{\frac{k}{2}(1+f)}{\rightleftharpoons \atop{\small \frac{k_{-}}{2}(1+f)}}
  L_1 + Q, & S + P \stackrel{\frac{k}{2}(1-f)}{\rightleftharpoons \atop{\small \frac{k_{-}}{2}(1-f) }}
  L_1 + P \\
  \\
  S \stackrel{\epsilon}{\rightleftharpoons \atop{\small \epsilon_{-} }} R_1, &
  S + P \stackrel{\frac{k}{2}(1+f)}{\rightleftharpoons \atop{\small \frac{k_{-}}{2}(1+f)}}
  R_1 + P, & S + Q \stackrel{\frac{k}{2}(1-f)}{\rightleftharpoons \atop{\small \frac{k_{-}}{2}(1-f)}}
  R_1 + Q.
\end{array}
\end{equation}
Here
\begin{equation}
Q = \sum_{n=1}^N n L_n, \qquad P = \sum_{n=1}^N n R_n,
\end{equation}
represent a measure of the total concentrations of left-handed and
right handed polymers. The amount of each chiral monomer produced is influenced
by the total amount of chiral oligomer already present in the system. We allow for the monomers
themselves to participate in these substrate reactions: hence $N
\geq n \geq 1$, where $N$ is the maximum chain length. The central
top and bottom reactions in Eq. (\ref{precursor}) are
enantioselective, whereas those on the right hand side are
non-enantioselective.  The model therefore contains the features of
limited enantioselectivity, first proposed as an alternative to the
mutual inhibition of Frank \cite{AG}.

An important observation is that differences in the Gibbs free energy
$\Delta G^0$ between initial and final states should be the same in
all the reactions listed in Eq.(\ref{precursor}), which implies the
thermodynamic constraint on the following forward and reverse reaction constants (see
also \cite{SH})
\begin{equation}
\frac{\epsilon}{\epsilon_{-}} = \frac{k}{k_{-}}.
\end{equation}
The polymerization and chain end-termination reactions (see below) are not subject to a thermodynamic constraint.

The monomers combine to form chirally pure polymer chains denoted by
$L_n$ and $R_n$, according to the isodesmic \cite{Meijer} stepwise
reactions for $1\leq n \leq N-1$
\begin{equation}\label{polybuild}
\begin{array}{cc}
  L_n + L_1 \stackrel{k_{aa}}{\rightleftharpoons \atop{\small k^{-}_{aa} }} L_{n+1}, &
  R_n + R_1 \stackrel{k_{bb}}{\rightleftharpoons \atop{\small k^{-}_{bb} }}
  R_{n+1},
\end{array}
\end{equation}
and inhibition, or the chain end-termination reactions for $N-1 \geq
n \geq 2$:
\begin{equation}\label{spoiledbuild}
\begin{array}{cc}
  L_n + R_1 \stackrel{k_{ab}}{\rightleftharpoons \atop{\small k^{-}_{ab} }} RL_{n}, &
  R_n + L_1 \stackrel{k_{ba}}{\rightleftharpoons \atop{\small k^{-}_{ba} }}
  LR_{n}.
\end{array}
\end{equation}
These upper limits for $n$ specified in
Eqs.(\ref{polybuild},\ref{spoiledbuild}) ensure that the
\textit{maximum} length for all oligomers produced (or consumed) by these reaction
sets, both the homo- and heterochiral ones, is never greater than
$N$. In the remainder of this paper we will consider here the natural and chiral symmetric rate assignments $k_{aa}=k_{bb}$, $k_{ab}=k_{ba}$ and likewise for the inverse
rates, $k^{-}_{aa}=k^{-}_{bb}$ and $k^{-}_{ab}=k^{-}_{ba}$. There are then four independent rate constants.

We include a separate reaction for
the heterodimer formation and dissociation:
\begin{equation}\label{heterodimer}
L_1 + R_1 \stackrel{k_{h}}{\rightleftharpoons \atop{\small k^{-}_{h} }} H \equiv L_1R_1,
\end{equation}
where $k_h=(k_{ab}+k_{ba})/2$ and $k^{-}_h = (k^{-}_{ab} + k^{-}_{ba})/2$. Note that these
latter two rate constants are automatically determined from the above choice and that
$L_1R_1 = R_1L_1$ of course.  This completes
the specification of the model's reactions.

The model is left-right symmetric, that is, possesses a discrete $Z_2$ symmetry \cite{Chaikin}, which is manifest in the
elementary reaction steps, in the rate constants, and in the corresponding differential rate equations (see below).
This symmetry can be broken spontaneously by the dynamical solutions of the differential rate equations.
This model is thus apt for investigating spontaneous mirror
symmetry breaking. Though not considered here, the effects of explicit chiral bias (e.g., that induced by
external physical fields) can also be studied
with this model by lifting the $Z_2$
degeneracy in the reaction rates, e.g., allowing for $k_{aa} \neq k_{bb}$, etc., leading to a maximum of eight
independent rate constants for describing the
reaction set in Eqs. (\ref{polybuild},\ref{spoiledbuild}).

Rate-equation theory as employed in chemical kinetics is used to describe the differential
rate equations of the achiral source, the monomers, as well as the homo- and heterochiral oligomers
belonging to this reaction network. The kinetic equations for the concentrations that follow from
these elementary steps are as follows. We begin with the rate equations for the
achiral precursor $S$ and those corresponding to the two chiral monomers:
\begin{eqnarray}\label{kinS}
\frac{d [S]}{dt} &=& -2\epsilon [S]-k [S](P+Q)
+ \epsilon_{-} ([L_1] + [R_1]) +\frac{1}{2}k_{-}
[L_1]\big((1+f)Q +
(1-f)P \big)\nonumber \\
&+& \frac{1}{2}k_{-} [R_1]\big((1+f)P + (1-f)Q\big),\\
\label{kinL1} \frac{d [L_1]}{dt} &=& \epsilon [S] + \frac{k}{2}
[S]\big((1+f)Q +
(1-f)P\big)
- k_{aa}[L_1]\Big(2[L_1] + \sum_{n=2}^{N-1}[L_n] + k_{ba}/k_{aa}
\sum_{n=1}^{N-1}[R_n] \Big)\nonumber \\
&-& \epsilon_{-} [L_1] - \frac{k_{-}}{2} [L_1]\big((1+f)Q +
(1-f)P\big)\nonumber \\
&+& k_{aa}^{-} \Big(2[L_2] + \sum_{n=2}^{N-1} [L_{n+1}] + k_{ba}^-/k_{aa}^-
\sum_{n=2}^{N-1} [LR_n] \Big)
+ k_h^- H,\\
\label{kinR1} \frac{d [R_1]}{dt} &=& \epsilon [S] + \frac{k}{2}
[S]\big((1+f)P +
(1-f)Q\big)
- k_{bb}[R_1]\Big(2[R_1] + \sum_{n=2}^{N-1}[R_n] + k_{ab}/k_{bb}
\sum_{n=1}^{N-1}[L_n] \Big)\nonumber
\\
&-&\epsilon_{-} [R_1] - \frac{k_{-}}{2} [R_1]\big((1+f)P +
(1-f)Q\big)\nonumber \\
&+& k_{bb}^- \Big(2[R_2] + \sum_{n=2}^{N-1} [R_{n+1}] + k_{ab}^{-}/k_{bb}^{-}
\sum_{n=2}^{N-1} [RL_n] \Big)
+ k_h^- H.
\end{eqnarray}
Whereas for $N-1 \geq n\geq 2$ we have the set of stepwise polymerization rate equations
\begin{eqnarray}\label{polyL}
\frac{d [L_n]}{dt} &=& k_{aa} [L_1]([L_{n-1}] - [L_n]) - k_{ab} [L_n][R_1]
+ k_{ab}^{-} [RL_n] + k_{aa}^{-}([L_{n+1}]-[L_n]),\\
\label{polyR}
\frac{d [R_n]}{dt} &=& k_{bb} [R_1]([R_{n-1}] - [R_n]) - k_{ba} [R_n][L_1]
+ k_{ba}^{-} [LR_n] + k_{bb}^{-}([R_{n+1}]-[R_n]).
\end{eqnarray}
Note, in accord with Eqs.(\ref{polybuild},\ref{spoiledbuild}) for
the largest polymers $n=N$, we have instead the final pair
\begin{eqnarray}\label{polyLN}
\frac{d [L_N]}{dt} &=& k_{aa} [L_1][L_{N-1}] -  k_{aa}^{-} [L_N],\\
\label{polyRN} \frac{d [R_N]}{dt} &=& k_{bb} [R_1][R_{N-1}] -  k_{bb}^{-}
[R_N].
\end{eqnarray}
Then the kinetic equation for the heterodimer $H \equiv L_1R_1$
(which we keep separate from the other end-chain rate equations):
\begin{equation}
\frac{d [H]}{dt} = k_h [L_1][R_1] - k_h^{-} [H].
\end{equation}

Lastly, for $N-1 \geq n \geq 2$ the rate equations for the ``end-spoiled" chains:
\begin{eqnarray}\label{spoilL}
\frac{d [LR_n]}{dt} &=& k_{ba} [L_1][R_n] -  k_{ba}^{-} [LR_n],
\\ \label{spoilR}
\frac{d [RL_n]}{dt} &=& k_{ab} [R_1][L_n] - k_{ab}^{-} [RL_n].
\end{eqnarray}
For chemical systems closed to matter flow, the constant mass
constraint that must be obeyed by the coupled system of differential
equations at every instant is given by (the overdot denotes time derivative):
\begin{eqnarray}\label{constantmasshet}
[\dot S] &+& 2[\dot H] + \sum_{n=1}^N n([\dot L_n] + [\dot R_n]) +
\sum_{n=2}^{N-1} (n+1)\big([\dot RL_n] + [\dot LR_n]\big) = 0.
\end{eqnarray}
This relation can be verified directly using the above set of $4N-2$ kinetic
equations Eqs. (\ref{kinS}-\ref{spoilR}).

\section{\label{sec:entropy} Entropy production}

The entropy production rate in an irreversible process is a measure
of the dissipation in that process. At equilibrium, the entropy
production rate vanishes and is an extremum \cite{RossVlad}. This
production has been investigated recently for reversible
versions of the Frank model \cite{KondeKapcha,Mauksch}. In those simple models, the behavior
of the entropy produced near the chiral symmetry breaking transition
as well as its subsequent temporal development,
depends on whether the chemical system is open or closed
to matter flow. We will return to this important point below. Here we
consider the behavior of the entropy produced by polymerization
reactions and monomer catalysis when the system undergoes a chiral phase
transition as well at the later stages when the system reaches final chemical
equilibrium upon racemization.

For reactions obeying mass action kinetics, the entropy produced in
any chemical reaction can be calculated straightforwardly in terms
of the individual elementary reaction rates \cite{PrigoKonde,RossVlad}. The
rate of entropy production is the sum over the difference of the
forward ($R_{jf}$) and reverse ($R_{jr}$) reaction rates multiplied by
the natural logarithm of the ratio of the forward and reverse rates
\cite{PrigoKonde}:
\begin{equation}\label{entropy}
\sigma(t) = R^* \sum_j (R_{jf}-R_{jr})\ln
\Big(\frac{R_{jf}}{R_{jr}}\Big) \geq 0,
\end{equation}
where the sum runs over each elementary reaction step $j$, and
$R^* = 8.314 \,{\rm J\, mol^{-1}\,
K^{-1}}$ is the gas constant. Since our reaction scheme is set up for
closed systems, equilibrium is reached after a racemization
time scale $t_{racem}$ is reached, which suggests a further measure is provided
by the total net entropy produced from the start of the
reactions through chiral symmetry breaking, then on to final racemization,
when the system reaches chemical and thermodynamic equilibrium and
$\sigma(t > t_{racem}) = 0$
\begin{equation}\label{totentropy}
\sigma_T = \int_0^{t_{racem}} \sigma(u) \, du.
\end{equation}
This quantifies the total dissipation over the complete history of
the chemical transformations under study.

The sum in Eq.(\ref{entropy}) contains $2N+4$ terms. In order
to determine which specific steps of the full reaction network
provide the leading contributions to $\sigma$, we
group the forward and reverse reaction rates as follows:

\subsection{\label{sec:entropy1} Direct monomer production}

%
\begin{eqnarray}
R_{1f}=\epsilon [S] &\qquad& R_{1r}=\epsilon_{-} [L_1],\\
R_{2f}=\epsilon [S] &\qquad& R_{2r}=\epsilon_{-} [R_1].
\end{eqnarray}
%

\subsection{\label{sec:entropy2} Monomer catalysis}

%
\begin{eqnarray}
R_{3f}=\frac{k}{2}(1+f) [S][Q] &\qquad& R_{3r}=\frac{k_{-}}{2}(1+f) [L_1][Q],\\
R_{4f}=\frac{k}{2}(1+f) [S][P] &\qquad& R_{4r}=\frac{k_{-}}{2}(1+f) [R_1][P],
\end{eqnarray}
\begin{eqnarray}
R_{5f}=\frac{k}{2}(1-f) [S][P] &\qquad& R_{5r}=\frac{k_{-}}{2}(1-f) [L_1][P],\\
R_{6f}=\frac{k}{2}(1-f) [S][Q] &\qquad& R_{6r}=
\frac{k_{-}}{2}(1-f) [R_1][Q].
\end{eqnarray}
%

\subsection{\label{sec:entropy3} Polymerization}

For $1 \leq n \leq N-1$ we have
\begin{eqnarray}
R_{nf}^L = k_{aa} [L_1][L_n] &\qquad& R_{nr}^L = k_{aa}^{-} [L_{n+1}], \\
R_{nf}^R = k_{bb} [R_1][R_n] &\qquad& R_{nr}^R = k_{bb}^{-} [R_{n+1}].
\end{eqnarray}
%

\subsection{\label{sec:entropy4} End-chain termination and the heterodimer}

The heterodimer rates are
\begin{equation}
R^h_f = k_h [L_1][R_1] \qquad R^h_r = k_h^{-} [H],
\end{equation}
whereas for $2 \leq n \leq N-1$ the end-chain forward and reverse
rates are given by
\begin{eqnarray}
R_{nf}^{eR} = k_{ab} [L_n][R_1] &\qquad& R_{nr}^{eR} = k_{ab}^{-} [RL_n], \\
R_{nf}^{eL} = k_{ba} [R_n][L_1] &\qquad& R_{nr}^{eL} = k_{ba}^{-}
[LR_n].
\end{eqnarray}
%

\section{\label{sec:results} Results}

As discussed in the Introduction, we are interested in testing the model's ability to amplify the initial
small statistical deviations about the idealized racemic composition \cite{Mislow}, in systems closed to matter flow and
taking microscopic reversibility fully into account.
In order to study the sensitivity of the above reversible polymerization network
to these minuscule initial enantiomeric excesses, a very dilute initial concentration of a scalemic (non racemic) mixture
was employed in the calculations: the initial monomeric concentrations of
$[L_1]_0 = (1 \times 10^{-6} + 1\times 10^{-15}) M$  and $[R_1]_0 = 1 \times 10^{-6} M$ yielding
an initial chiral excess of $ee_0 = 5 \times 10^{-8}\%$. This is actually slightly lower \cite{Mills,Mislow} than the
excess corresponding to the initial monomer concentrations ($ee_0 = 6.1 \times 10^{-8}\%$).
The initial concentration of the achiral substrate is $[S]_0 = 2M$, whereas
those corresponding the homo- hetero-oligomers are all set to zero:
$[L_n]_0=[R_n]_0 = 0$, for $2 \leq n \leq N$, $[H]_0 = 0$, and $[LR_n]_0 = [RL_n]_0 = 0$ for $2 \leq n \leq N-1$.
We choose $\epsilon = 2\times 10^{-5}$, $\epsilon_{-}= 10^{-10}$, $k=2.0$, $k_{-}=10^{-5}$, $f=0.9$,
$k_{aa}=k_{bb}= 1.0$,
$k_{ab}=k_{ba}= 10^3$, $k_{ab}^{-}=k_{ba}^{-}= 1.0$,
$k_{aa}^{-}=k_{bb}^{-}= 10^{-5}$. For illustrative purposes, we consider chains that can grow to a maximum length of $N=12$. This
$N$ value is intended as a mean ``ball-park" figure suggested by recent experiments yielding homochiral chains anywhere from $N=5$
\cite{Hitzb} up to
$N=18$ \cite{Lahavb} chiral repeat units, depending on the amino acids employed and the experimental conditions.
The differential rate equations Eqs. (\ref{kinS}-\ref{spoilR}) were numerically integrated with the version 7 Mathematica  program
package and using a level of numerical precision, typically twenty or more significant digits, to ensure the
numerical significance of these initial concentrations and enantiomeric excess.
The results were monitored and verified to assure that total system mass Eq. (\ref{constantmasshet}) remained constant in
time.

Results are quantified in terms of a variety of convenient chiral measures.
The percent enantiomeric excess values of the oligomers with homochiral sequence are
calculated according to ($2\leq n \leq N)$
\begin{equation}\label{een}
ee_n = \frac{[L_n] - [R_n]}{[L_n] + [R_n]}\times 100.
\end{equation}
A global or ensemble-averaged measure of the degree of symmetry breaking is provided by
the number-weighted enantiomeric excess $\eta$:
\begin{equation}\label{averagedEE}
\eta = \frac{\sum_{n=2}^N([L_n] - [R_n])}{\sum_{n=2}^N([L_n] +
[R_n])}\times 100.
\end{equation}
The importance of the enantiomeric excess is that it provides the order parameter for the symmetry breaking transition:
the $|ee| \geq 0$ is zero for chiral symmetric states and nonzero otherwise. In the latter case, the $Z_2$ symmetry is broken.
The average chain length of the homopolymers is given by:
\begin{equation}\label{nbar}
{\bar n} = \frac{\sum_{n=2}^N n([L_n] + [R_n])}{\sum_{n=2}^N ([L_n] +
[R_n])},
\end{equation}
and the root mean square deviation in the homochiral chain length are:
\begin{equation}\label{standard}
(\overline{n^2})^{1/2} \equiv \sqrt{<(n- {\bar n})^2>} = \sqrt{<n^2> - <n>^2}.
\end{equation}
The angular brackets $< >$ denote averaging with respect to the ensemble $\sum_n ([L_n]+[R_n])$, similar to
Eq.(\ref{nbar}).
It is important to remember that these are all time-dependent quantities.

\begin{figure}[h]
\includegraphics[width=0.50\textwidth]{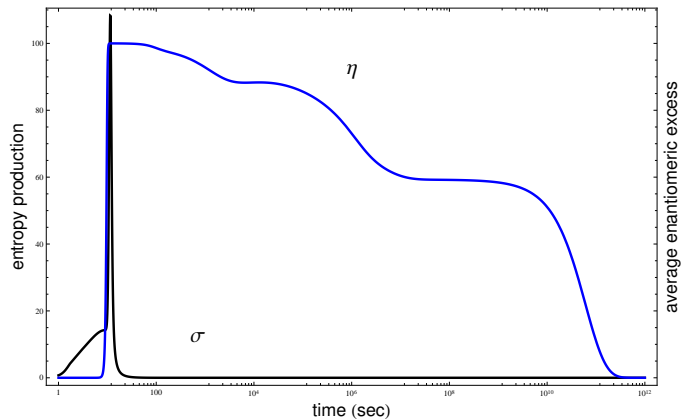}
\caption{\label{entropyprod} Time evolution (logarithmic time scale) of the average enantiomeric excess $\eta$,
averaged over all chains ($2 \leq n \leq 12$) Eq. (\ref{averagedEE}), and the associated entropy production
$\sigma$, Eq. (\ref{entropy}). The entropy production peaks sharply at the onset of chiral symmetry breaking followed by a dramatic decrease to
very small values. $\sigma$ strictly approaches zero only at the racemization time scale $t_{racem} \gtrsim 10^{11}$.
Initial concentrations: $[L_1]_0 = (1 \times 10^{-6} + 1\times 10^{-15}) M$  and $[R_1]_0 = 1 \times 10^{-6} M$ ($ee_0 = 5 \times 10^{-8}\%$),
$[S]_0 = 2M$, all other homo- and heterochiral oligomer initial concentrations are zero;  and for the following rates $\epsilon = 2\times 10^{-5}$, $\epsilon_{-}= 10^{-10}$, $k=2.0$, $k_{-}=10^{-5}$, $f=0.9$,
$k_{aa}=k_{bb}= 1.0$,
$k_{ab}=k_{ba}= 10^3$, $k_{ab}^{-}=k_{ba}^{-}= 1.0$,
$k_{aa}^{-}=k_{bb}^{-}= 10^{-5}$.}
\end{figure}

\begin{figure}[h]
\includegraphics[width=0.50\textwidth]{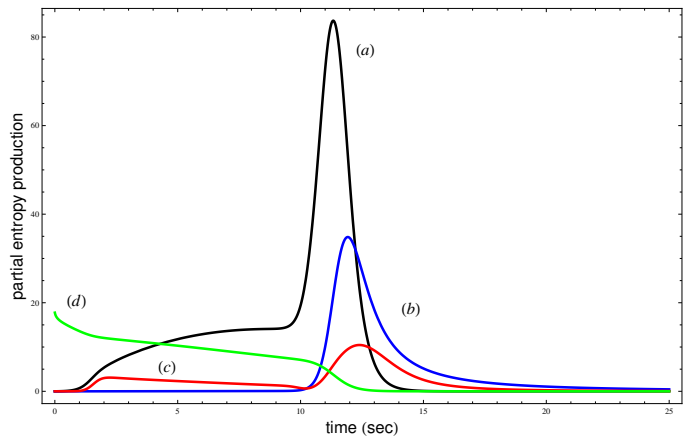}
\caption{\label{partialentropies} Partial contributions to the total entropy production. From, (a) limited enantioselective monomer catalysis,
(b) stepwise polymerization reactions, (c) chain-end termination and heterodimer formation reactions $\times 20$, and
(d) direct monomer production $\times 1000$. Note how the peak values of (a), (b) and (c) are displaced in time.
Same initial concentrations and rate constants as in Fig. \ref{entropyprod}.}
\end{figure}

Temporary but rather long lived asymmetric amplification can take place, as shown in
Figure \ref{entropyprod}; note the logarithmic time scale.  The enantiomeric excess $\eta$ averaged over all chain lengths Eq. (\ref{averagedEE}), from the dimer
on up to the maximum length chain $N=12$ starts off at zero value until a time on the order of $t\sim 10$
at which the excess increases rapidly to nearly $100\%$ at SMSB. This is followed by a gradual stepwise decrease
or chiral erosion characterized by
the appearance of quasi-plateaus of approximate constancy: $\eta$ falls to about $90 \%$ at $t \sim 10^3$, then to about $60 \%$
at $t \sim 10^6$, staying approximately level until the final decrease to zero occurring at a time
on the order of $t \sim 10^{11} - 10^{12}$. The system has racemized on this time scale.
No appreciable differences in $\eta$ can be discerned when we include the monomer, that is, start the
sums at $n=1$: we still observe slow chiral erosion proceeding though quasi-steady plateaus.
The rate of the entropy produced Eq. (\ref{entropy}) exhibits an initial increase followed by a
dramatic burst coinciding with the mirror symmetry breaking transition.
This production then decreases rapidly to an exceedingly tiny but
non zero value that remains constant during the entire period of slow chiral erosion, spanning more than ten orders of magnitude in
time. The entropy production then goes strictly to zero when the system racemizes in complete accord with the fact that the
system has reached chemical equilibrium.
Although not displayed in Figure \ref{entropyprod}, the substrate concentration falls from its initial value to zero
approximately coincident with the peak structure of the entropy production, thus suggesting a connection between the sharp
production of the latter and the change in $S$.
The total entropy produced Eq.(\ref{totentropy})
is $\sigma_T = 378.4 \,{\rm J\, mol^{-1}\,
K^{-1}}$.
We can identify the major contributions to the entropy production in this process, see Figure \ref{partialentropies}, from
calculations of the partial entropies. In this way we find that the leading contribution comes from the monomer catalysis steps Eq.(\ref{precursor}), followed by
the polymerization itself Eq.(\ref{polybuild}), next by the mutual inhibition reactions: heterodimer formation Eq.(\ref{heterodimer})
and the ``end-spoiled" cross-inhibition reactions Eq.(\ref{spoiledbuild}).
The least important contribution comes from the direct production of monomers from the achiral substrate. The first three partial contributions all
display a peak structure in the neighborhood of the symmetry breaking transition, with the corresponding peak values being displaced
in time, see Figure \ref{partialentropies}. The exception to this is the entropy rate due to direct monomer production, which decreases monotonically.


\begin{figure}[h]
\includegraphics[width=0.50\textwidth]{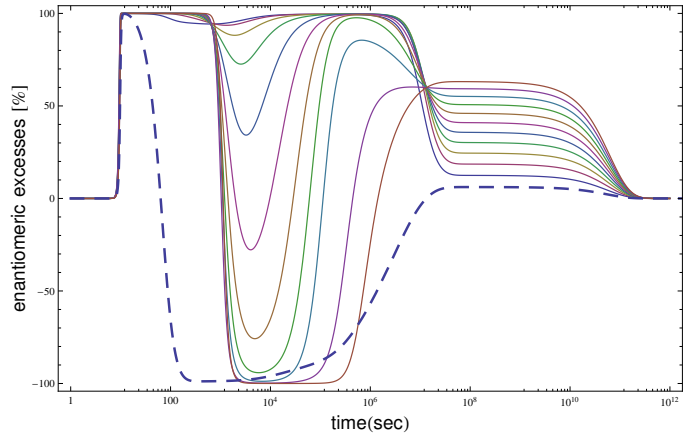}
\caption{\label{allees} Time dependence (logarithmic scale) of the individual chain-length
dependent enantiomeric excesses $ee_n \%= \frac{[L_n] - [R_n]}{[L_n] + [R_n]}\times 100$, from the start of reactions to
chiral symmetry breaking, and then on to the final racemization (family of solid curves). Near
$t\simeq 10^4$, the sequence of curves from top to bottom
corresponds to $n=2$ to $n=12$, respectively.
Note damped oscillatory behavior of the excesses corresponding to $n=7,8,9,10,11,12$.
The dashed curve shows the chiral excess for the monomers: $ee_{1}\%$.
Same initial concentrations and rate constants as in Fig. \ref{entropyprod}.}
\end{figure}

\begin{figure}[h]
\begin{center}
\begin{tabular}{cc}
\includegraphics[width=0.49\textwidth]{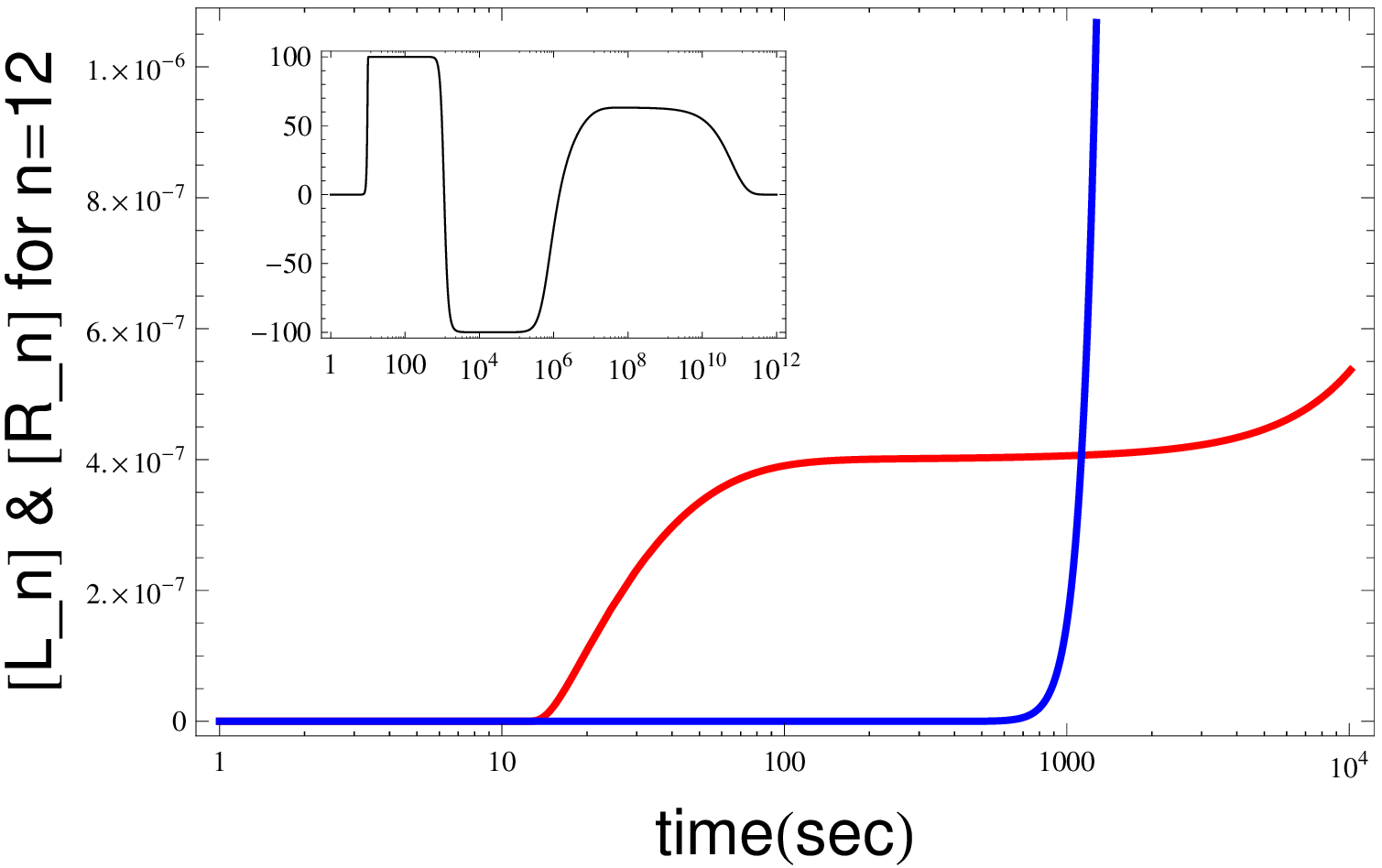}&
\includegraphics[width=0.47\textwidth]{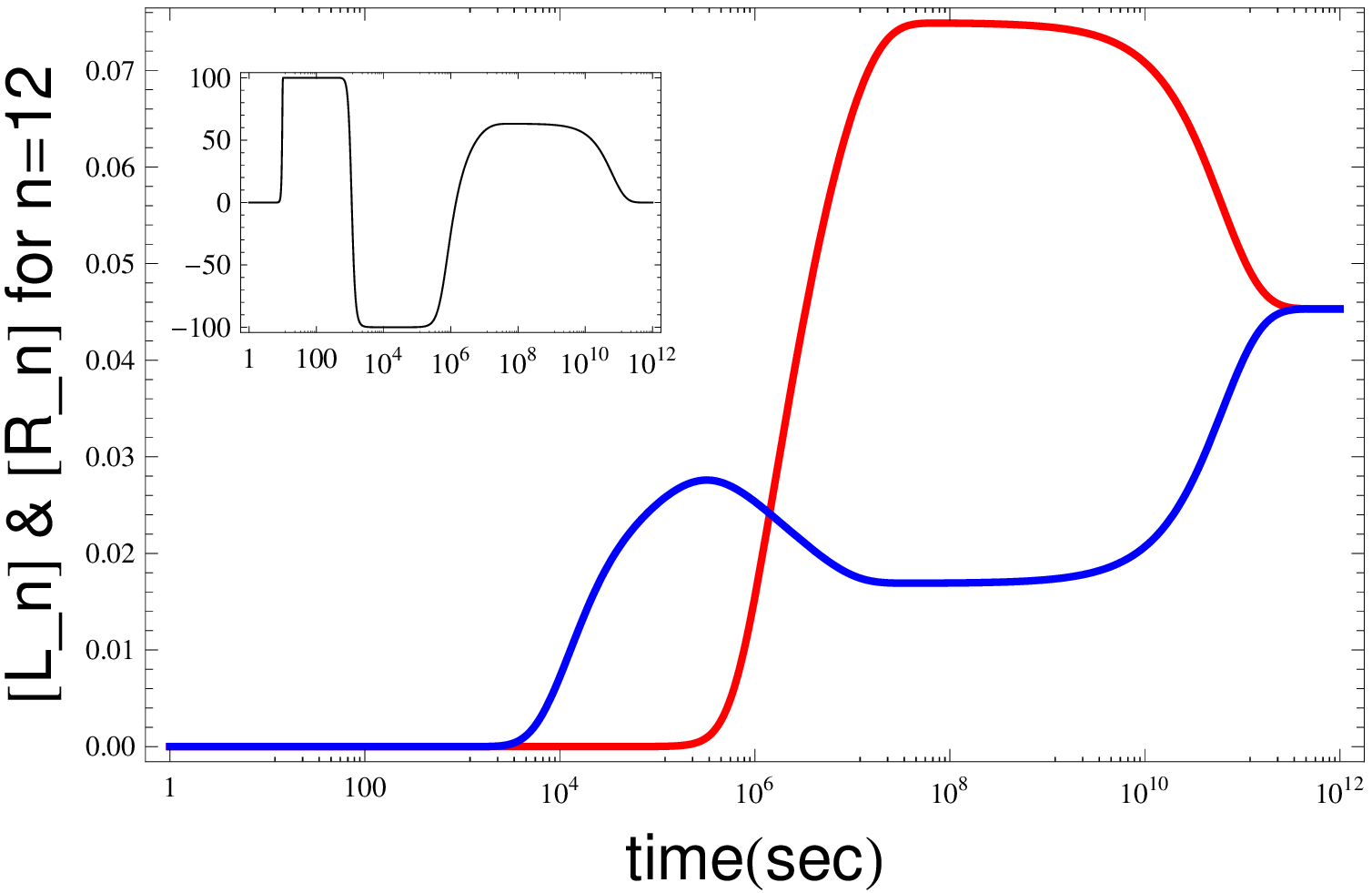}\\
\end{tabular}
\caption{\label{oscillations} Chiral oscillations. Evolution of the individual concentrations $[L_{12}]$
(the upper curve at the bifurcation) and $[R_{12}]$ for the complete time interval spanning symmetry breaking to
final racemization.  The inset graph gives the associated percent enantiomeric
excess $ee_{12}$ versus time. Left: expanded view of the initial stages of evolution for $1 \leq t \lesssim 10^4$. Symmetry breaking occurs at $t \sim 10$ with $L_{12}$ dominating the excess.
The chiral excess of these longest chains vanishes instantaneously for the \textit{first} time at around
$t=10^3$ when the curves intersect, and then turns over such that now $R_{12}$ dominates the chiral excess, see right hand graph (this leads to the sign flip in the excess, see inset).
Right: chiral excess vanishes a \textit{second} time at $t \sim 10^6$ when the two curves intersect again (compare to inset). Then from about $10^6$ to $10^{11}$ the $L_{12}$ chains
again dominate the chiral excess until racemization, when the two curves collapse to a common curve (anti-bifurcation).
Same initial concentrations and rate constants as in Fig. \ref{entropyprod}.}
\end{center}
\end{figure}

\begin{figure}[h]
\begin{center}
\begin{tabular}{cc}
\includegraphics[width=0.48\textwidth]{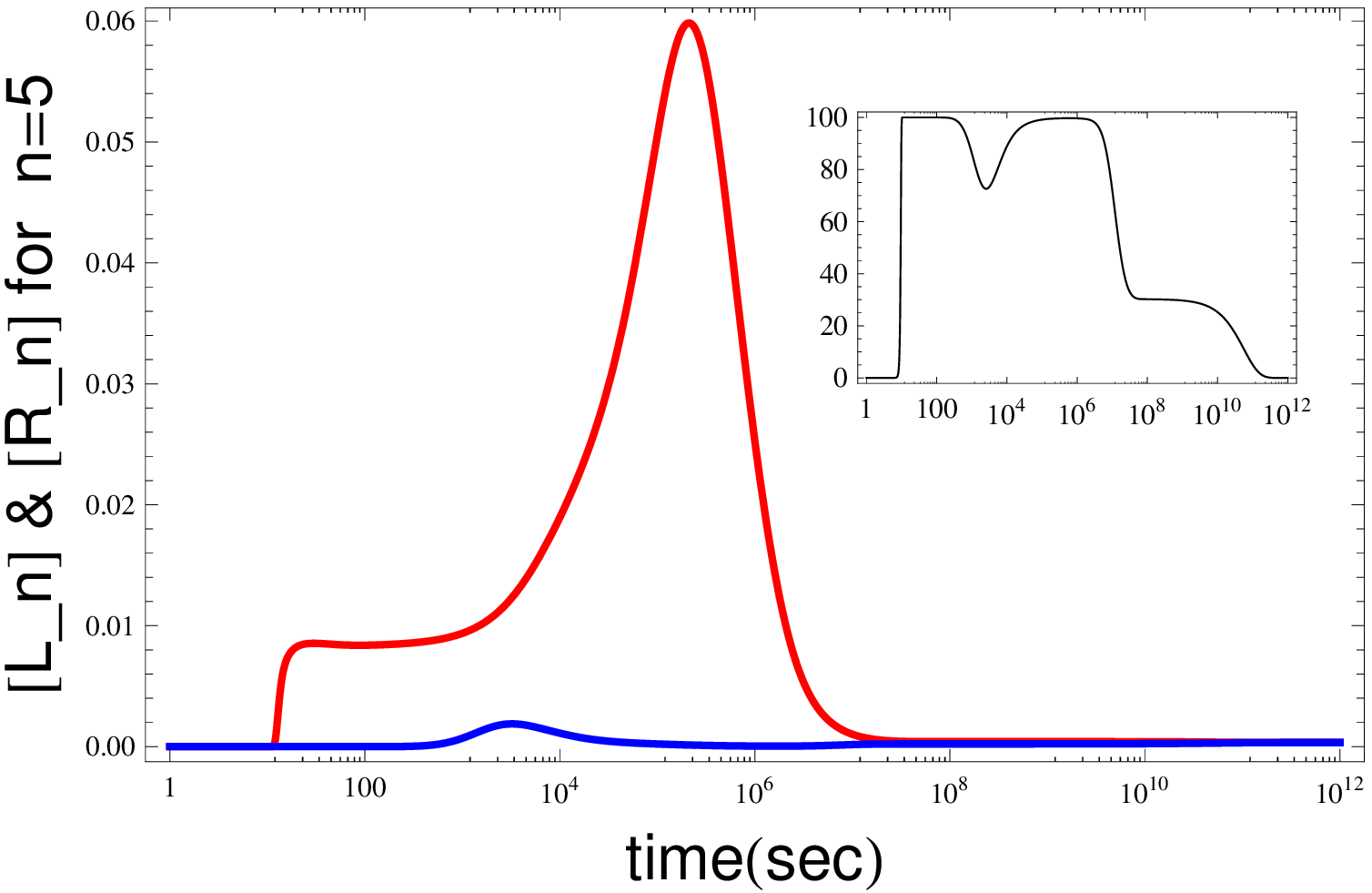}&
\includegraphics[width=0.49\textwidth]{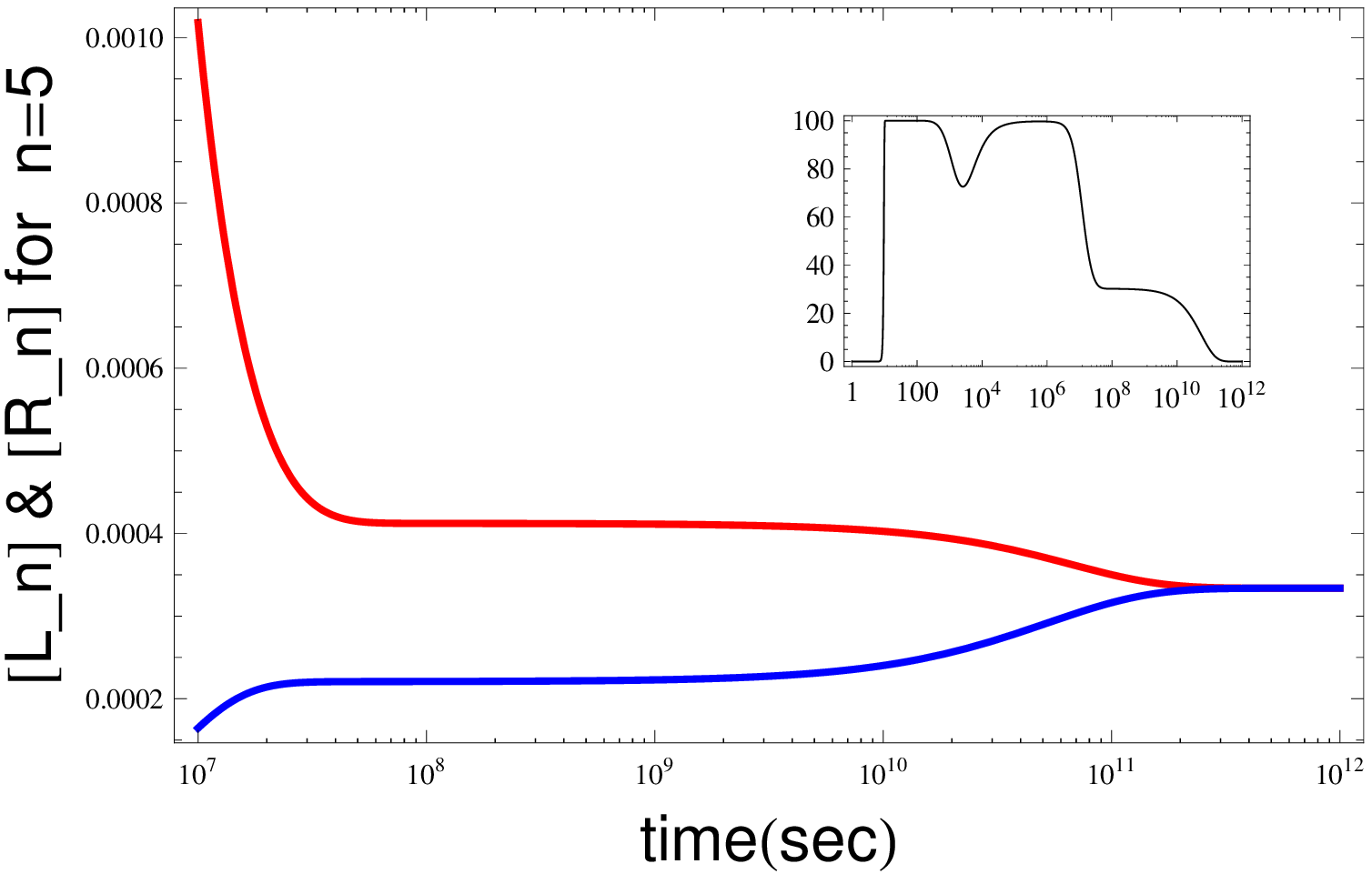}\\
\end{tabular}
\caption{\label{pentamers} Evolution of the individual concentrations $[L_{5}]$ (the upper curve at the bifurcation) and $[R_{5}]$
for the complete time interval spanning symmetry breaking to
final racemization.  The inset graph gives the associated percent enantiomeric
excess $ee_{5}$ as a function of time. Left: full history of the evolution. Right: close up of the final time scales and
racemization (anti-bifurcation).}
\end{center}
\end{figure}

A finer or more detailed measure of the degree of symmetry breaking and amplification is provided by the individual
percent chain-length dependent enantiomeric excesses, Eq.(\ref{een}). A remarkable and complex dynamic behavior is revealed here.
The time dependence of these $n$-dependent $ee$'s is plotted
in Figure \ref{allees}; note the logarithmic time scale. The individual $ee$'s follow a common curve from initialization to
chiral symmetry breaking, at about $t \sim 10$, and remain together at nearly $100 \%$ until about $t \sim 100$ at which time the common curve
begins to split up into its
constituents. Then, the percent chiral excess of each length homochiral chain behaves differently, until they again coalesce into a single curve upon final racemization,
occurring at around $t \sim 10^{11}$. There is a common tendency for all the $ee$'s to decrease at intermediate time scales, with the largest length chains
(here, $N=7,8,9,10,11,12$) passing from \textit{positive} then to \textit{negative} values
of the excess. The $N=12,11,10$ chains exhibit nearly $-100\%$ excess during the period from $10^3$ to $10^4$ and beyond: there has been a chiral \textit{sign reversal}
in the excess corresponding to the largest chains. This holds also for the monomer $ee_1$, plotted in the dashed curve.
Except for the monomer, the individual excesses then all increase back to positive values at $t \sim 10^6$, then from $t \sim 10^7$ to
$t \sim 10^{11}$, the excess increases sequentially as a  function of the chain length $n$ until racemization, where they all collapse to zero.
The temporal behavior of the enantiomeric excesses of the largest chains $n=12,11,10,9,8,7$ is reminiscent of strongly \textit{damped oscillations}.
This oscillatory behavior in the enantiomeric excess can be understood in terms of the evolution of the individual concentrations of the longest
chains. To illustrate this, we focus on the
time dependence of the concentrations $[L_{12}]$ and $[R_{12}]$, the corresponding $ee$ experiences the largest amplitude damped oscillations, see Figure \ref{oscillations}.
For reference the inset diagram shows the
enantiomeric excess over the entire time interval of the simulation, compare to Figure \ref{allees}. As shown in Figure \ref{oscillations}, the dominant concentration
shifts back and forth between $[L_{12}]$ and $[R_{12}]$, respectively, until racemization when both concentrations converge to a common value.
This leads to the chiral oscillations depicted in the inset. The shorter chains do not experience this oscillation, as illustrated for example by the
concentrations $[L_{5}]$ and $[R_{5}]$ plotted in Figure \ref{pentamers}. There, the dominant concentration is always $[L_5]$, all the way from symmetry breaking at
$t \sim 10$  to
racemization, at approximately $t \gtrsim 10^{11}$. The corresponding $ee$ suffers a dip near $t =10^3$ (see inset), due to the concentration $[R_5]$ momentarily
increasing at that time, see left hand graph of Figure \ref{pentamers}. This dip becomes more pronounced the longer the chain, see the sequence of curves
around $t \sim 10^4$ in Figure \ref{allees}, and becomes a
fully-fledged damped chiral oscillation for the longest chains in the system.

\begin{figure}[h]
\begin{center}
\begin{tabular}{cccc}
\includegraphics[width=0.24\textwidth]{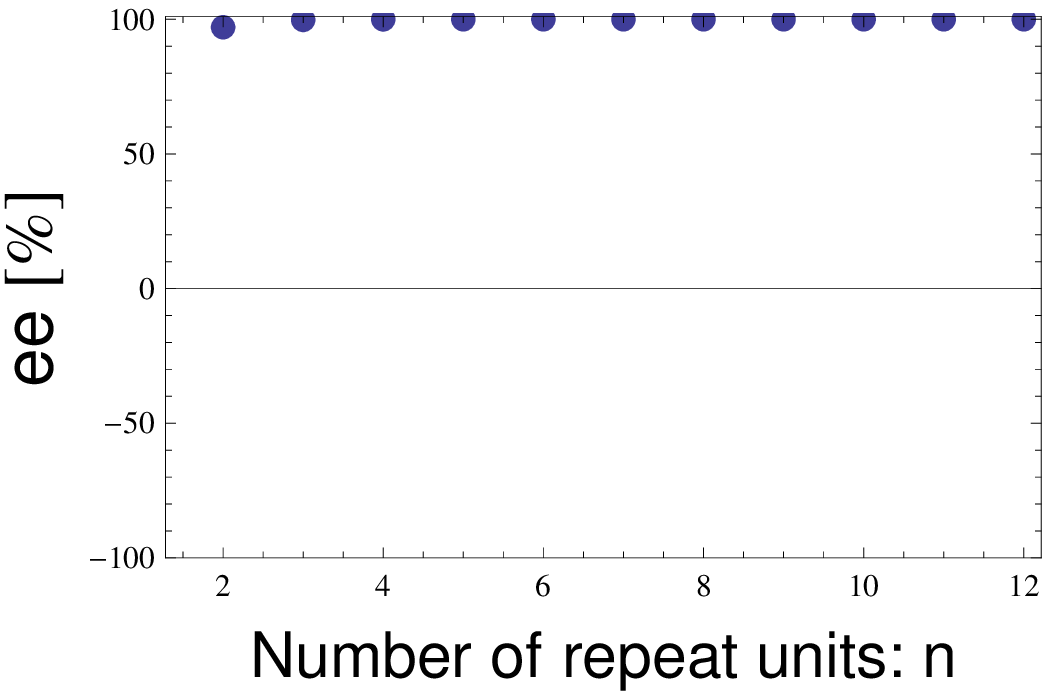}&
\includegraphics[width=0.24\textwidth]{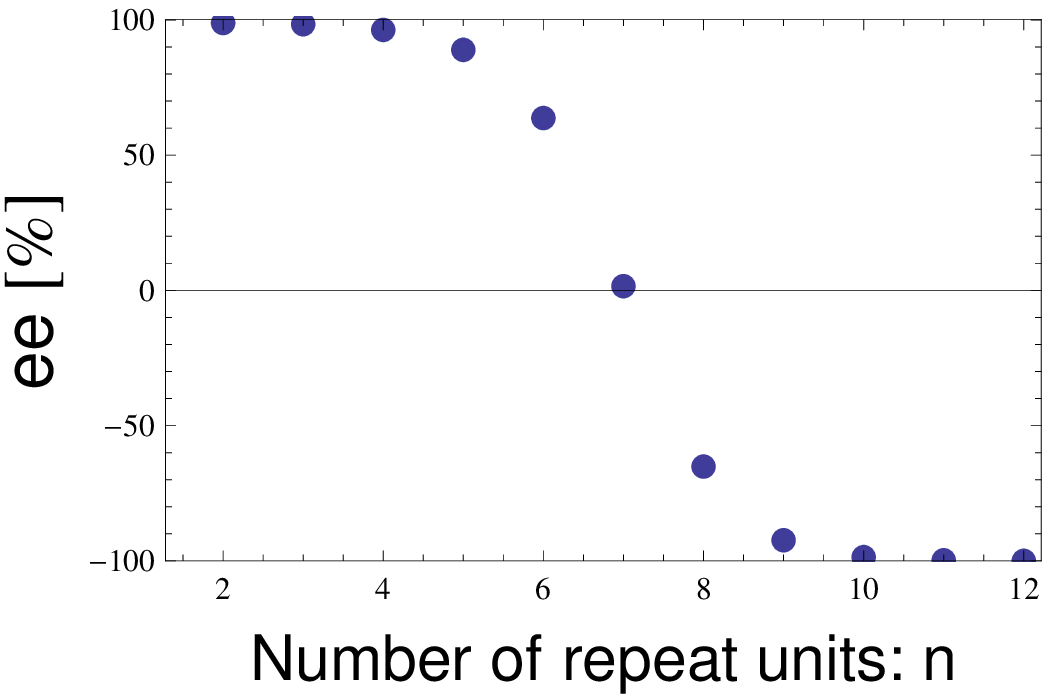}&
\includegraphics[width=0.24\textwidth]{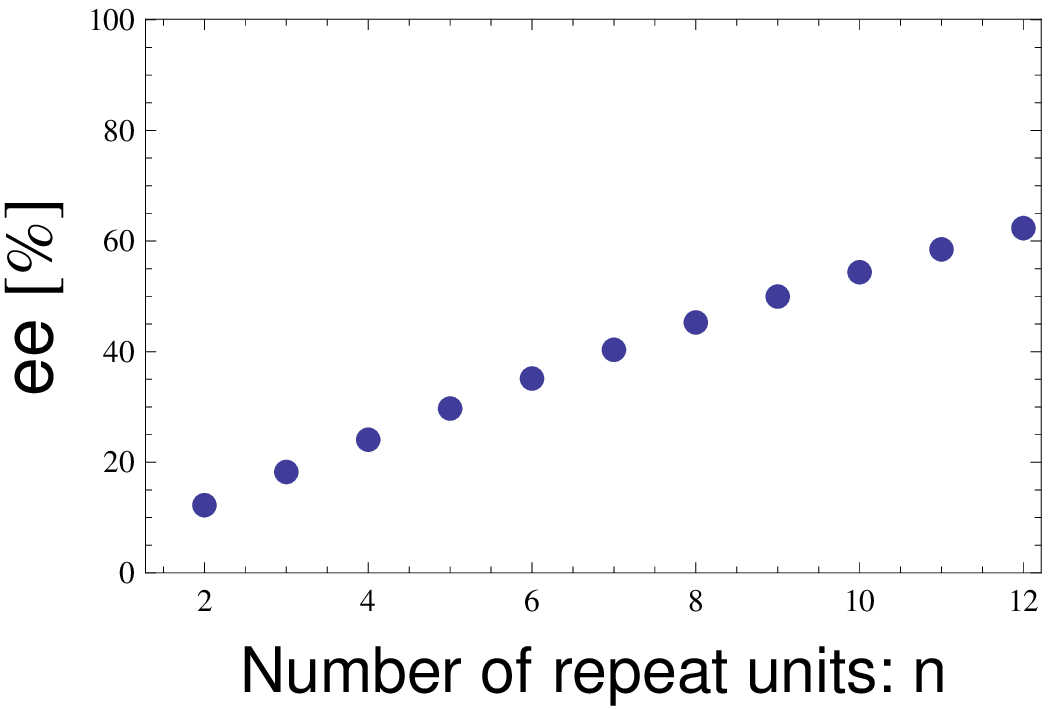}&
\includegraphics[width=0.24\textwidth]{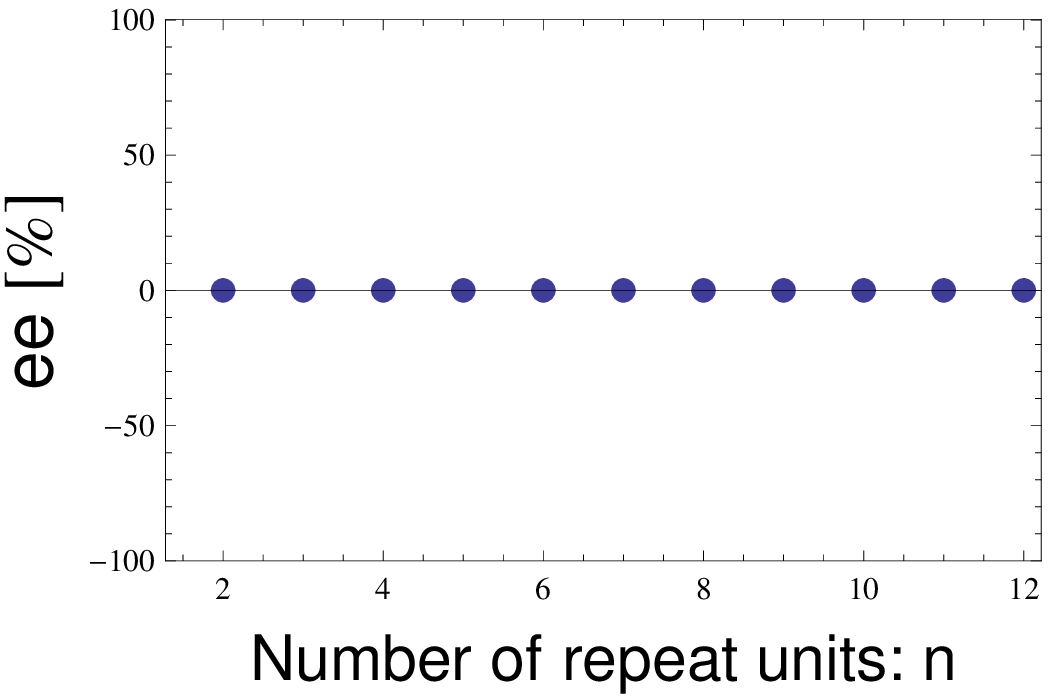}\\
\end{tabular}
\caption{\label{ees} Different time slices or ``snap-shots" of Fig \ref{allees} showing dependence of the chain-length
dependent enantiomeric excesses $ee_n \%= \frac{[L_n] - [R_n]}{[L_n] + [R_n]}\times 100$ for $n\geq 2$, at different selected time scales.
From left to right: total chiral symmetry breaking for all length homopolymer chains at $t=100$, next, second graph
shows the sign reversal tendency for the largest chains at $t=10^4$, followed by the third graph, the monotone increase of chiral excess
as a function if chain length at
$t=10^9$, and then the fourth graph, the final racemization at $t=10^{13}$. Same initial concentrations and rate constants as in Fig. \ref{entropyprod}.}
\end{center}
\end{figure}

\begin{figure}[h]
\includegraphics[width=0.50\textwidth]{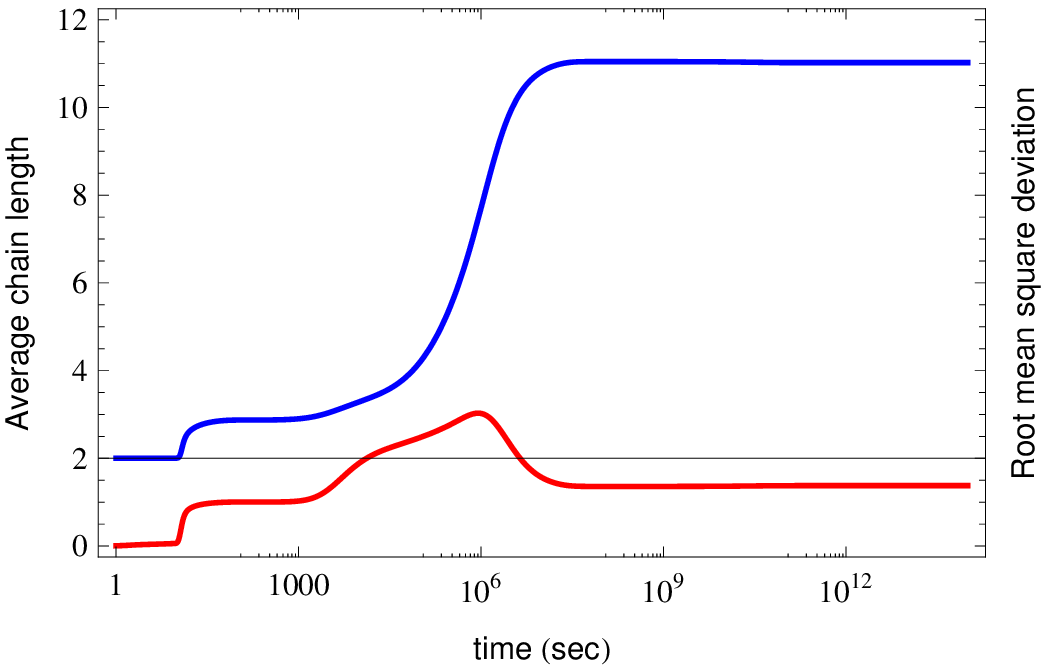}
\caption{\label{chainlength} The time evolution (logarithmic scale) of the average or mean homopolymer chain length ${\bar n}$ (upper curve)
and the corresponding root-mean-square deviation $<\overline{n^2}>^{\frac{1}{2}}$ from the mean value (lower curve).
The final stable values of the mean
and RMS values are ${\bar n} = 11.02$ and $<\overline{n^2}>^{\frac{1}{2}} = 1.38$, for $t \gtrsim 10^{7}$.
This demonstrates that the final racemic mixture is dominated by the longer length homopolymer chains, and this is the
final equilibrium configuration. Same initial concentrations and rate constants as in Fig. \ref{entropyprod}.}
\end{figure}
\begin{figure}[h]
\includegraphics[width=0.50\textwidth]{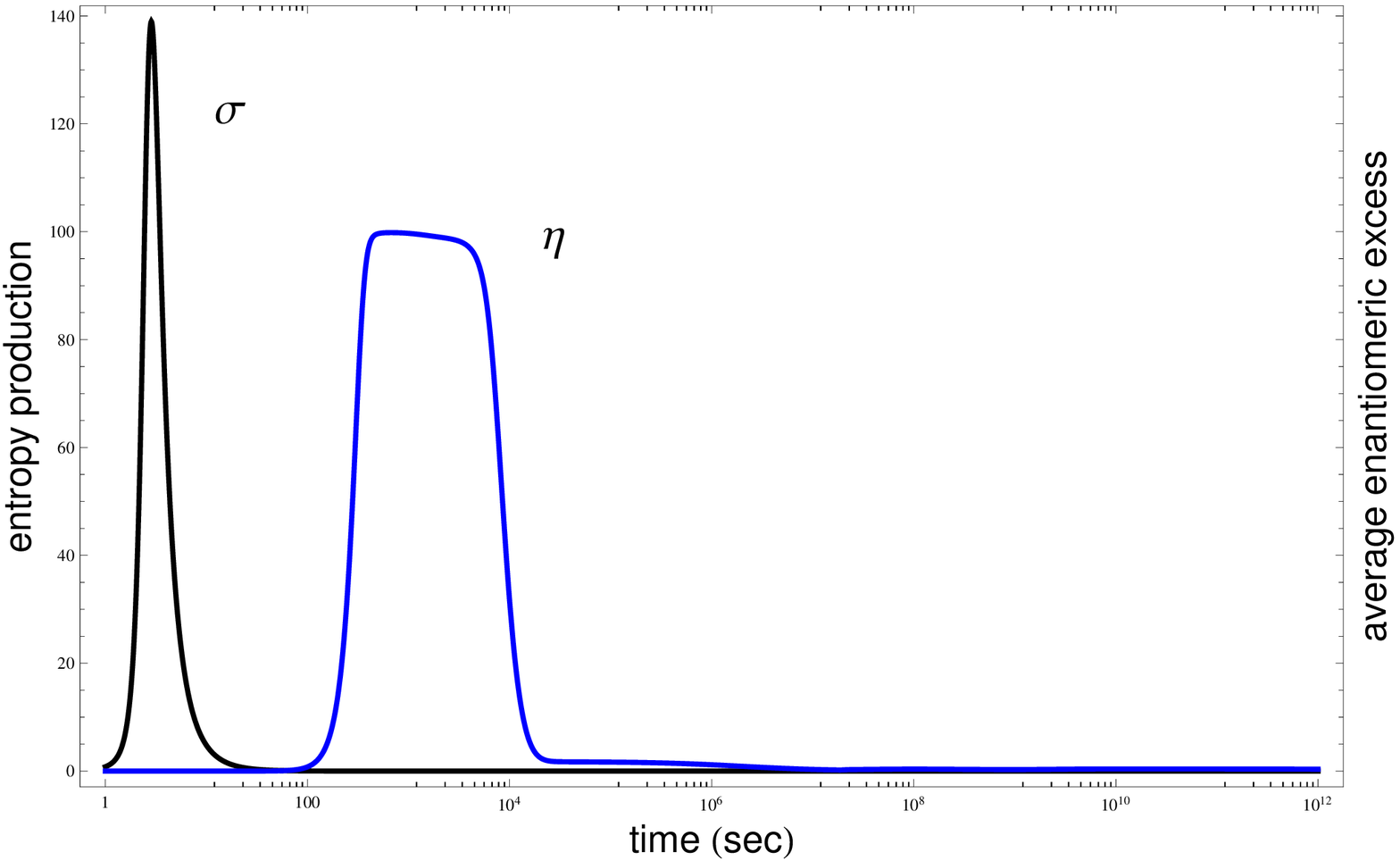}
\caption{\label{entropyprod2} Time evolution (logarithmic time scale) of the average enantiomeric excess $\eta$,
averaged over all chains ($2 \leq n \leq 12$) Eq. (\ref{averagedEE}), and the associated entropy production
$\sigma$, Eq. (\ref{entropy}). The entropy production peaks well before ($t\sim 3$) the onset of chiral symmetry breaking
($t \sim 300$) followed by a dramatic decrease to
very small values. $\sigma$ strictly approaches zero only at the racemization time scale $t_{racem} \gtrsim 10^{15}$.
Initial concentrations: $[L_1]_0 = (1 \times 10^{-6} + 1\times 10^{-15}) M$  and $[R_1]_0 = 1 \times 10^{-6} M$ ($ee_0 = 5 \times 10^{-8}\%$),
$[S]_0 = 2M$, all other homo- and heterochiral oligomer initial concentrations are zero;  and for the following rates $\epsilon = 2\times 10^{-5}$,
$\epsilon_{-}= 10^{-10}$, $k=2.0$, $k_{-}=10^{-5}$, $f=0.9$,
$k_{aa}=k_{bb}= 1.0$,
$k_{ab}=k_{ba}= 20$, $k_{ab}^{-}= k_{ba}^{-}= k_{aa}^{-}= k_{bb}^{-}= 10^{-6}$}
\end{figure}
\begin{figure}[h]
\includegraphics[width=0.50\textwidth]{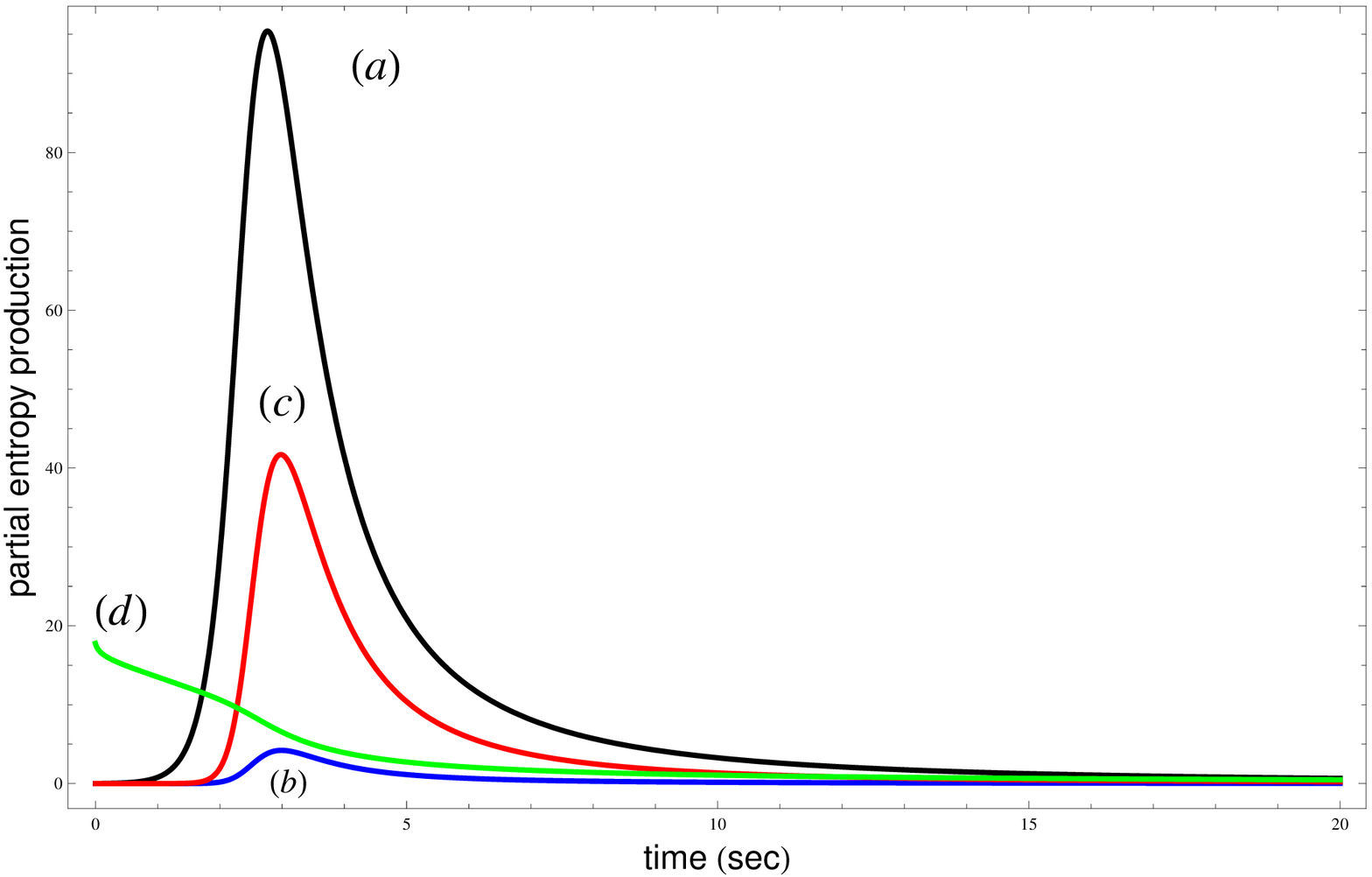}
\caption{\label{partialentropies2} Partial contributions to the total entropy production. From, (a) limited enantioselective monomer catalysis,
(b) stepwise polymerization reactions, (c) chain-end termination and heterodimer formation reactions $\times 20$, and
(d) direct monomer production $\times 1000$. Same initial concentrations and rate constants as in Fig. \ref{entropyprod2}.}
\end{figure}
\begin{figure}[h]
\begin{center}
\begin{tabular}{cc}
\includegraphics[width=0.45\textwidth]{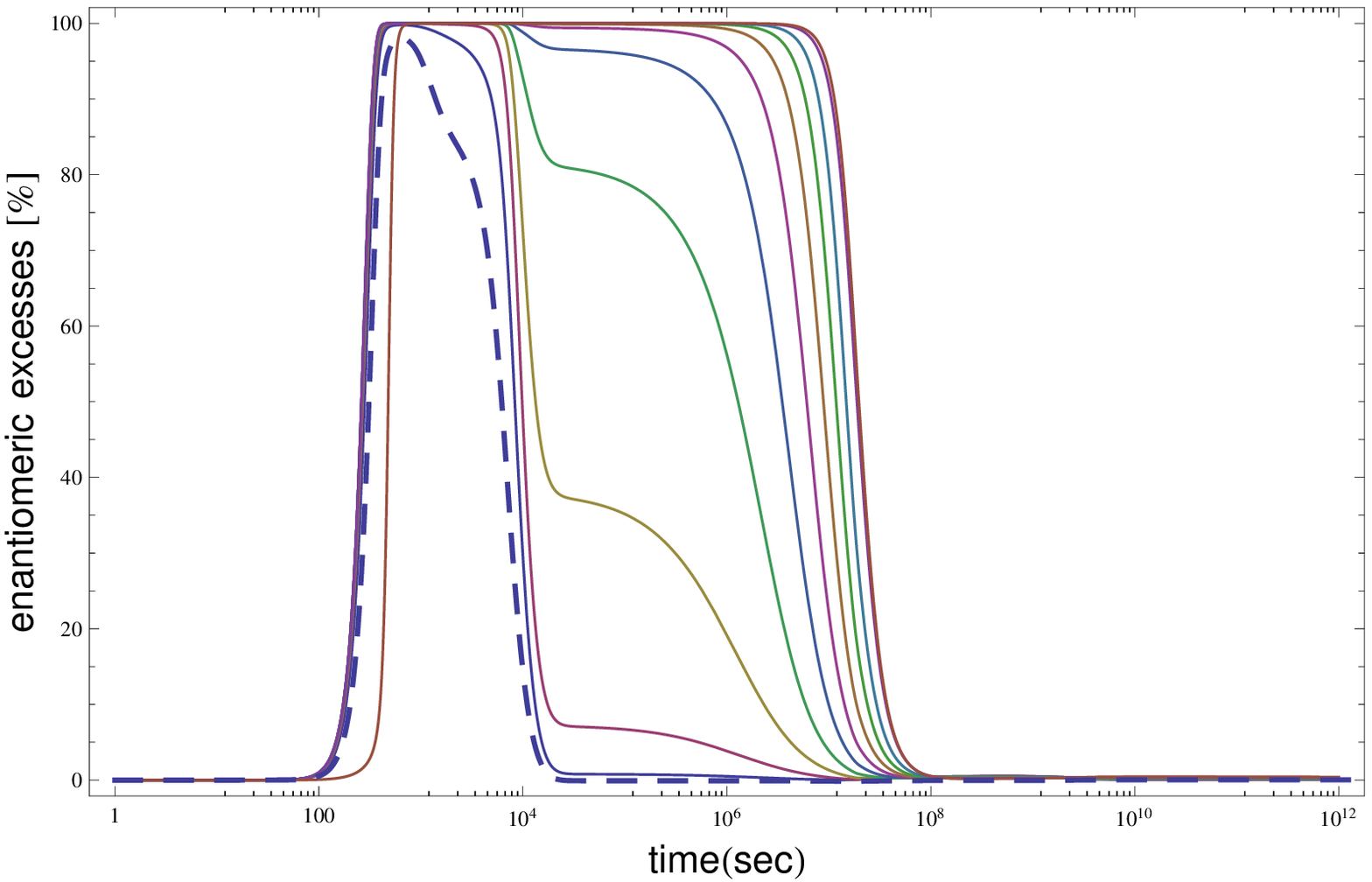}&
\includegraphics[width=0.45\textwidth]{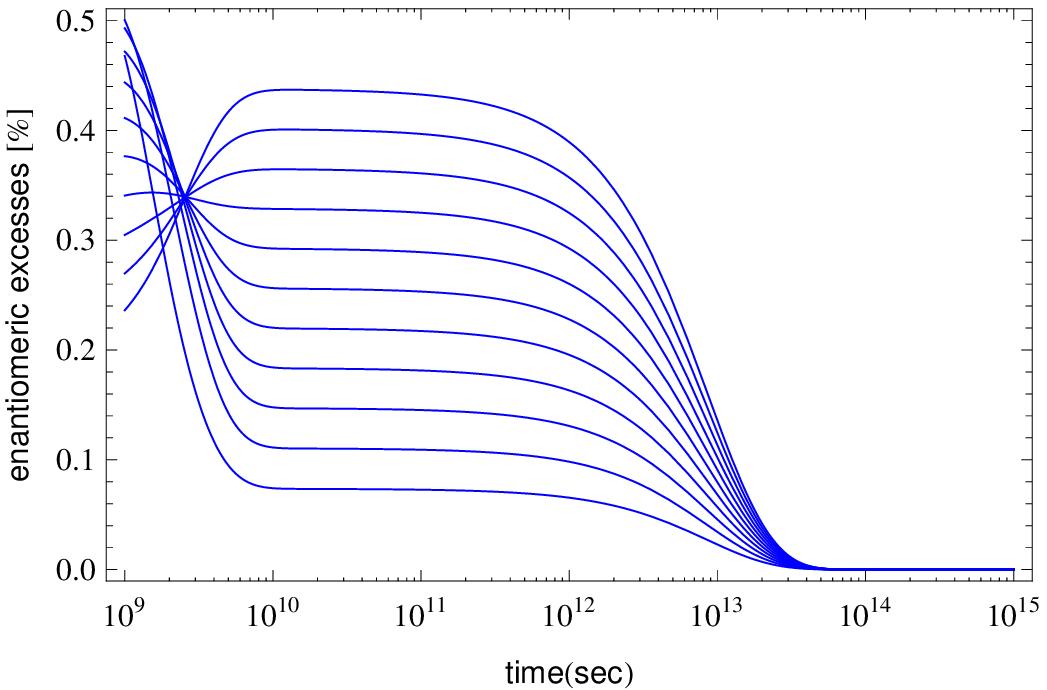}\\
\end{tabular}
\caption{\label{allees2} Time dependence (logarithmic scale) of the individual chain-length
dependent enantiomeric excesses $ee_n \%= \frac{[L_n] - [R_n]}{[L_n] + [R_n]}\times 100$, from the start of reactions to
chiral symmetry breaking, and then on to the final racemization (family of solid curves).  The dashed curve shows the chiral excess for
the monomers. Right hand side shows a blow-up of the $ee_n$'s for the time scale
$t = 10^9$ to $10^{15}$, exhibiting the sequence of excesses and its final convergence to zero at racemization.
Same initial concentrations and rate constants as in Fig. \ref{entropyprod2}.}
\end{center}
\end{figure}
\begin{figure}[h]
\begin{center}
\begin{tabular}{cccc}
\includegraphics[width=0.24\textwidth]{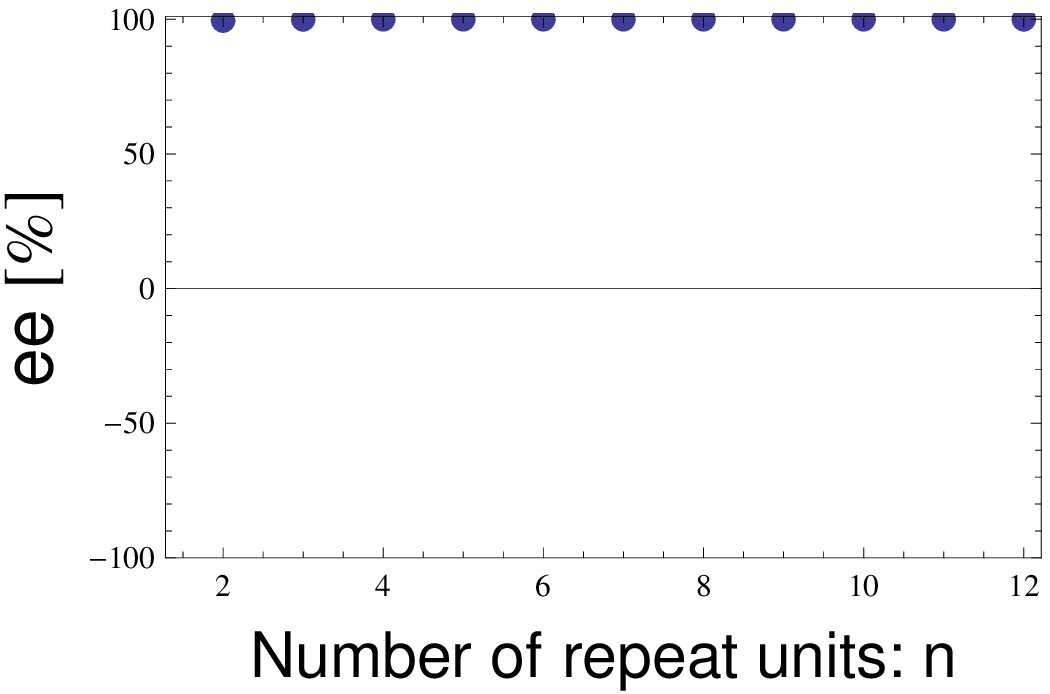}&
\includegraphics[width=0.24\textwidth]{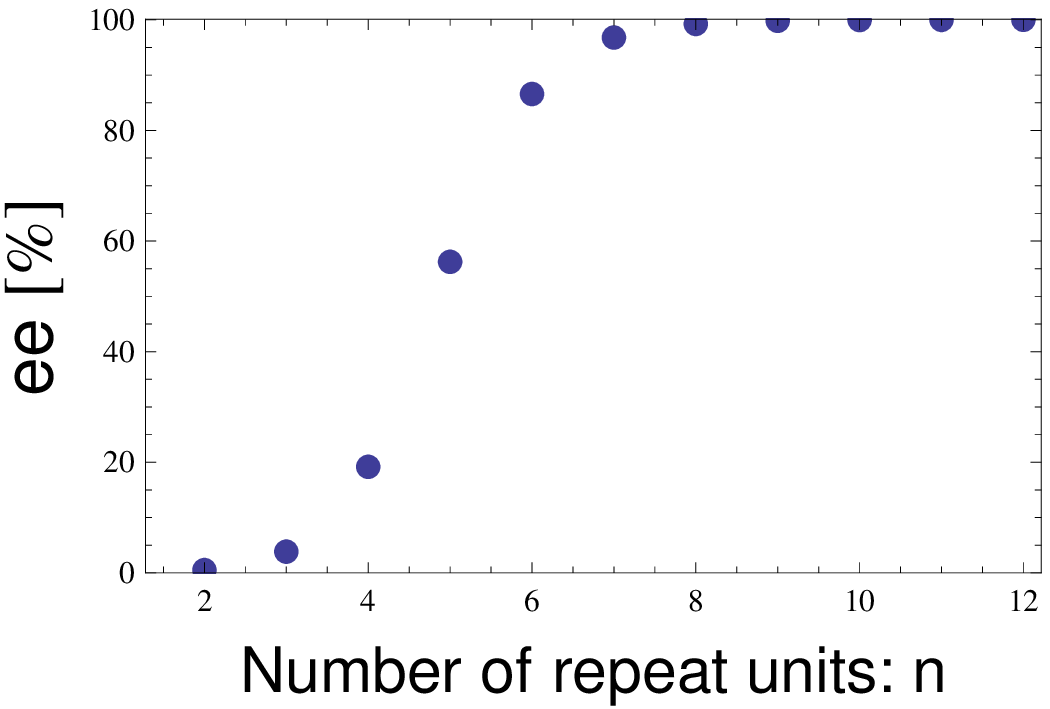}&
\includegraphics[width=0.24\textwidth]{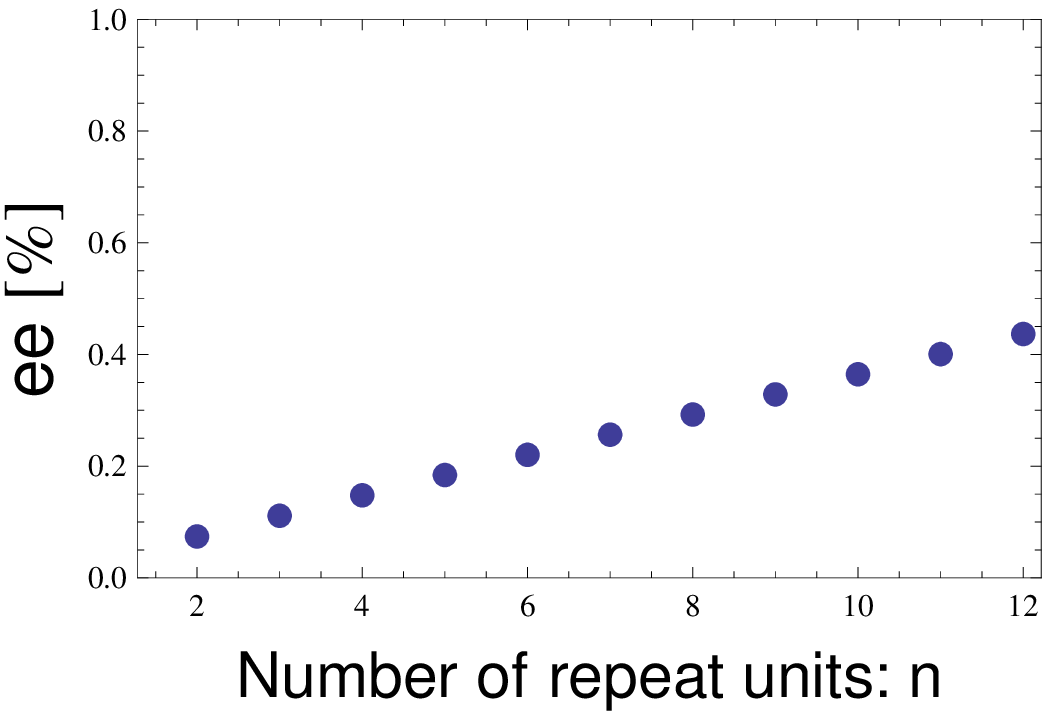}&
\includegraphics[width=0.24\textwidth]{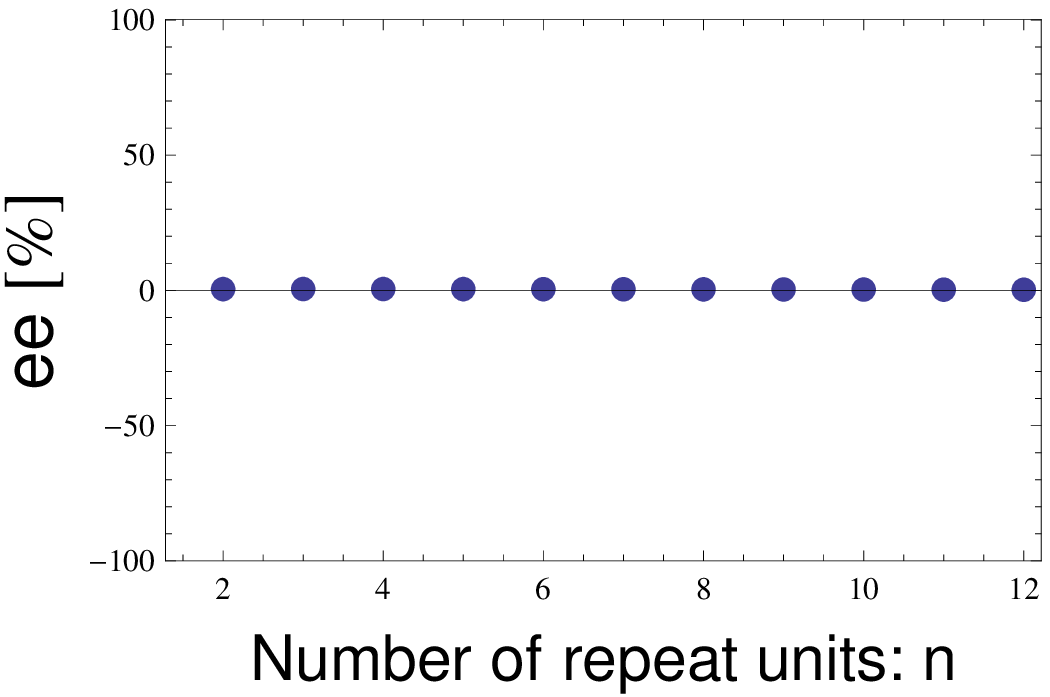}\\
\end{tabular}
\caption{\label{ees2} Different time slices or ``snap-shots" of Fig \ref{allees2} showing dependence of the chain-length
dependent enantiomeric excesses $ee_n \%= \frac{[L_n] - [R_n]}{[L_n] + [R_n]}\times 100$ for $n\geq 2$, at different selected time scales.
From left to right: total chiral symmetry breaking for all length homopolymer chains at $t=10^3$, next, second graph
shows the stepwise increase for the largest chains at $t=10^6$, followed by the third graph, the monotone increase of chiral excess
as a function if chain length at
$t=10^{10}$, and then fourth graph, the final racemization at approximately $t=10^{14}$. Same initial concentrations and rate constants as in Fig. \ref{entropyprod2}.}
\end{center}
\end{figure}

Static ``snapshots" of this dynamic behavior nicely complement the evolution of the chain length dependent enantiomeric excesses. In Figure \ref{ees}
we display the enantiomeric excess versus the number of chiral repeat units at selected time slices. In the
leftmost graph, the $ee$'s are all at $100\%$ for all the chains. The next graph, corresponding to $t = 10^4$, shows the sign reversal tendency as
a function of chain length, with the
full reversal ($-100\%$) being attained for the largest homochiral chains. The following graph, corresponding to $t = 10^9$ shows the monotone
increase of $ee$ with chain length. Finally, the righthand most graph shows that racemization has set in by $t = 10^{13}$.
These should be contrasted with Figure \ref{allees}. It is interesting to point out that both the qualitative behaviors depicted in the second and
third snapshot have been reported in two recent and independent polymerization experiments.
The tendency of the sign reversal in $ee$ (from positive to negative values) as a function of chain length has been observed
in the polymerization of racemic valine (Val-NCA) and leucine (Leu-NCA)in water subject to chiral initiators \cite{Lahava}.
By contrast, the monotonic increase of the percent $ee$ with chain length has been measured in independent chiral amplification experiments
using leucine and glycine in water \cite{Hitzb} starting with a $20\%$ initial enantiomeric excess of the $L_1$ monomer.
These static snapshots also raise the important question of \textit{when} to observe
the chiral amplification and the enantiomeric excesses. In nonlinear reaction schemes such as this one, the enantiomeric excesses one measures can depend
strongly on when the measurement or observation is made, that is, when one decides to \textit{terminate} the experiment.

Additional information regarding the homo-oligomer composition of the chemical system is provided by the average homochiral chain length $<n>$, see Eq.(\ref{nbar}).
We plot this in Figure \ref{chainlength} along with the standard deviation about the mean, Eq. (\ref{standard}).
The mean chain length starts off at $2$, corresponding to the homodimer and then increases monotonically after the
symmetry breaking transition, reaching a constant plateau at $t = 10^7$ where it remains constant all the way
through to racemization and beyond. The final mean value $<n> = 11.0$. The corresponding root mean square indicates that the
fluctuations in the mean chain length increase as the mean chain length increases but then drops down to a constant value
$(\overline{n^2})^{1/2} \sim 1.4$ when the average value stabilizes.
This indicates that the  final racemic composition is dominated by the longer chains: $n_{final} = 11.0 \pm 1.4$.
The racemization time scale depends on how ``irreversible" the model is. By way of example, if we increase the rate $k_{-}$ of
the reverse catalysis steps in Eq.(\ref{precursor}), keeping everything else constant, then the increased recycling of monomers back into
achiral precursor $S$ lowers this time scale as
follows: $(k_{-},t_{racem})= (10^{-6}, 5\times 10^{12} s),(10^{-5}, 5\times 10^{11} s), (10^{-4}, 5\times 10^{10} s), (10^{-3}, 1\times 10^{10} s),
(10^{-2}, 5\times 10^{9} s).$ By the same token, if we make $k_{-}$ smaller, we can postpone racemization.

The enantiomeric cross inhibition $k_{ab}=k_{ba}$ is a determining factor in this model.
By way of contrast, we consider a second $N=12$ run with a much lower mutual inhibition than employed above, namely
$k_{ab}=k_{ba}= 20$, and with the following inverse rates all set equal
$k_{aa}^{-}=k_{bb}^{-}= k_{ab}^{-}=k_{ba}^{-} = 10^{-6}$, but keeping the remainder of the rates
as before and with the same initial concentrations and excess. In this situation, the symmetry breaking occurs at a later time and
most interestingly, the entropy production now peaks well \textit{before} the mirror symmetry is broken, see Figure \ref{entropyprod2}.
Figure \ref{partialentropies2} shows that the catalysis still yields the major contribution to this peak, but the second and third most important contributions
are now formation of end-chain spoiled oligomers followed by the polymerization, exactly opposite to the previous run employing the much
higher mutual inhibition. The peak in $\sigma$ is due principally to monomer catalysis, and not symmetry breaking.

The time dependence of these $n$-dependent $ee$'s is plotted
in Figure \ref{allees2}; note the logarithmic time scale. The individual $ee$'s follow a common curve from initialization to
chiral symmetry breaking, at about $t \sim 300$, and remain together at nearly $100 \%$ until about $t \sim 1000$ at which time the common curve
begins to split up into its component parts. Note how the shorter homopolymers tend to racemize before the longer ones, there is a sequential
chiral erosion in the individual enantiomeric excesses that is more pronounced the shorter the chain length. This
holds as well for the monomer, plotted in the dashed curve (contrast to the monomer behavior in Fig. \ref{allees}).
Then, the percent excess of each length homochiral chain behaves differently, until they again coalesce into a single curve upon final racemization,
occurring at around $t \sim 10^{14}$. The final approach to racemization is qualitatively very similar to the case treated above, compare the
sequence in the right hand graph of Figure \ref{allees2}; to the sequence of curves in Figure \ref{allees} from roughly $t \sim 10^7$ to $10^{12}$.
A sequence of snap-shots of the $ee_n$'s
at selected times is displayed in Figure \ref{ees2}.

\begin{figure}[h]
\includegraphics[width=0.50\textwidth]{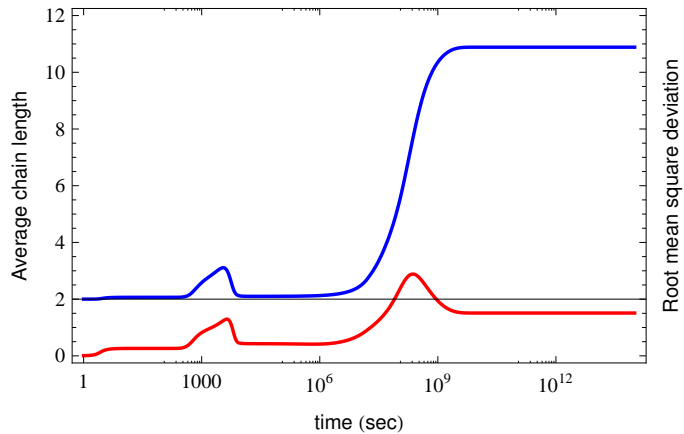}
\caption{\label{chainlength2} The time evolution (logarithmic scale) of the average or mean homopolymer chain length ${\bar n}$ (upper curve)
and the corresponding root-mean-square deviation $<\overline{n^2}>^{\frac{1}{2}}$ from the mean value (lower curve). The final stable values of the mean
and RMS values are ${\bar n} = 10.9$ and $<\overline{n^2}>^{\frac{1}{2}} = 1.5$, for $t \gtrsim 10^{10}$.
The final racemic mixture is highly dominated by the longer length homopolymer chains, and this is the
equilibrium configuration. Same initial concentrations and rate constants as in Fig. \ref{entropyprod2}.}
\end{figure}

Finally, we plot the average homochiral chain length $<n>$, see Eq. (\ref{nbar}) in Figure \ref{chainlength2} along with
the standard deviation. The mean chain length starts off at $2$, corresponding to the homodimer and then increases monotonically after the
symmetry breaking transition, reaching a constant plateau at about $t = 10^9$ where it remains constant all the way
through to racemization and beyond. The final mean value $<n> = 10.9$.
Once again, the  final racemic composition is dominated by the longest chains.
%

\section{\label{sec:conclusions} Conclusions and Discussion}

We have demonstrated that a strong chiral amplification can take place in
a reversible model of chiral polymerization closed to matter flow and subject to constraints imposed by micro-reversibility.
The inherent statistical fluctuations about the idealized racemic composition are
modeled by an initial minuscule enantiomeric excess in a system dilute in
the monomers. These results are important, because they suggest that spontaneous mirror symmetry breaking
in experimental chiral polymerization can take place, and with observable and large chiral excesses
without the need to introduce chiral initiators \cite{Lahavb} or large initial chiral excesses \cite{Hitzb}. Instead, the
needed chiral monomers (i.e., amino acids) can be produced directly from an achiral precursor and amplified via
catalysis. Strong mutual inhibition is required to amplify the initial $ee$ to large values, very similar
to what we found for the reversible Frank model in closed systems \cite{CHMR}. The chain-length
dependent enantiomeric excesses depend on time in a highly nontrivial way.
The essential rate constant is that corresponding to the enantiomeric inhibition.
A most intriguing novel feature revealed here for appreciable enantiomeric cross inhibition is the tendency for the chain length dependent enantiomeric excesses to exhibit
a damped oscillatory behavior before the onset of final racemization. In these conditions, the observed chiral excess is clearly
a time dependent phenomena, though the ``period" of the chiral oscillations can be quite long.
Oscillatory dynamics in chemical reactions has been observed experimentally, and analyzed theoretically and numerically in simple model
systems \cite{Scott,CH}; as far as we are aware, this
behavior has not been revealed previously as a valid dynamical solution in polymerization models. The implications for
chirality transmission are far reaching: ``memory" of the sign of the initial chiral fluctuation is washed-out by the oscillations in the
enantiomeric excess, adding another heretofore unexpected element of randomness to the process. While the sign of the
initial chiral fluctuation is entirely random, any subsequent chiral oscillations can further ``erase" the memory of the
sign of this initial enantiomeric excess. These oscillations cease as the system approaches its equilibrium state.
Moderate values leads to strong temporary symmetry breaking and larger values can lead to
long period damped chiral oscillations before final racemization takes over.

We have also shown that the rate of entropy production
per unit volume exhibits a peak value either before or near the vicinity of the
chiral symmetry breaking transition. This increase to a peak value is mainly due to the
catalytic production of the chiral monomers, followed next by the
stepwise polymerization reactions, and then by the chain-end termination reactions
and lastly, by the direct monomer production. The rate falls to a vanishingly small but constant nonzero
value maintained during the intermediate time scales, then drops to zero once the system has
racemized. Previous calculations of the entropy produced in chiral symmetry breaking transitions have been
carried out in the Frank model. In \cite{KondeKapcha}, $\sigma$ was evaluated for a reversible open flow Frank model
with a constant inflow of achiral substrate and a constant outflow of the mutual inhibition product.
In that situation, the entropy production rises from small initial value and then levels off to a
constant plateau after symmetry breaking. As the \textit{open flow} keeps this system far from equilibrium, it
can never racemize, and entropy is produced at a constant rate, \textit{in sempiternum}.
In \cite{Mauksch}, the rate of entropy
production was evaluated for a reversible open flow Frank model
including limited enantioselectivity. Mirror symmetry is broken incompletely, and $\sigma$ increases sharply
from the start of the reaction, remains at a constant level during the lifetime of the initial racemic
state and then decreases to a new stationary value once symmetry breaking sets in.
These latter authors also consider the reversible Frank model with constant concentration of substrate but freely varying inhibition product. The entropy
production increases, a peak value is reached when symmetry breaking is almost complete, then decreases to
very small values. On the other hand, for a reversible Frank model \textit{closed} to matter flow,
and with strong mutual inhibition, the rate of entropy production exhibits a sharp peak at the onset of symmetry breaking, falling to a tiny positive value until the
system racemizes, at which time $\sigma$ goes to zero \cite{private}. This latter behavior is qualitatively similar to
what we find in our polymerization model for large inhibition.

For sake of computational simplicity, we have considered a model wherein no generally mixed heteropolymers were formed, only
the heterodimer $LR=RL$ and $LR_{n-1}$ and $RL_{n-1}$ for $3 \leq n \leq N$.
The corresponding number of differential equations
grows linearly with the length $N$ as $4N-2$ . A more
realistic model should of course include the formation of all the possible heteropolymers of a given length $n$, i.e., those
heterochiral chains of length $n$ containing
$r \geq 1$  copies of $L_1$ and $s \geq 1$ copies of $R_1$ such that $r + s = n$. It is possible to build such a system, for example,
starting with the copolymerization model of \cite{WC2}, where the concentration variables are denoted as $c^L_{r,s}(t)|_{r \geq 1,s \geq 0}$
and $c^R_{r,s}(t)|_{r \geq 0,s \geq 1}$, the superscript
indicating the final monomer in the chain, while the double subscript $r,s$ encodes the number of individual $L$ and $R$'s
making up the chain, respectively. The number of differential equations in this case grows quadratically with polymer length $N$ as $N(N+1)$, and
is mathematically more involved. Detailed studies employing a \textit{reversible} copolymerization scheme in closed systems will
be presented elsewhere \cite{CdTH}.

\begin{acknowledgments}
We are grateful to Josep M. Rib\'{o} for a careful reading of the
manuscript and for numerous useful discussions and to Michael Stich for
his interest in oscillatory phenomena in chemical systems.
This research is supported in part by the Grant
AYA2009-13920-C02-01 from the Ministerio de Ciencia e Innovaci\'{o}n
(Spain) and forms part of the ESF COST Action CM07030: \textit{Systems
Chemistry}. CB is the recipient of a Calvo-Rod\'{e}s training and research
scholarship from the Instituto Nacional de T\'{e}cnica Aeroespacial
(INTA).
\end{acknowledgments}

\end{document}